\documentclass[%
 reprint,
superscriptaddress,
 amsmath,amssymb,
 aps,
 prl,
]{revtex4-2}

\usepackage{graphicx}
\usepackage{dcolumn}
\usepackage{bm}
\usepackage{miller}
\usepackage{setspace}
\usepackage{siunitx}
\usepackage{nicefrac}
\usepackage{enumerate}
\usepackage{multirow}
\usepackage{todonotes}
\usepackage{tabularx}
\usepackage{makecell} 
\usepackage{tablefootnote}
\usepackage{amssymb}
\usepackage{comment}
\usepackage{hyperref}  
\usepackage{xr}  
\usepackage{soul}
\usepackage{float}
\floatstyle{plaintop}
\restylefloat{table}
\usepackage{enumitem}

\begin{document}

\preprint{prl}

\title{Measuring vacancy-type defect density in monolayer semiconductors}

\author{Aleksandar Radi\'{c}}\email{ar2071@cam.ac.uk}
\affiliation{Cavendish Laboratory, Department of Physics, University of Cambridge, JJ Thomson Ave, Cambridge, United Kingdom}%

\author{Nick von Jeinsen}%
\affiliation{Cavendish Laboratory, Department of Physics, University of Cambridge, JJ Thomson Ave, Cambridge, United Kingdom}%

\author{Vivian Perez}
\affiliation{Cavendish Laboratory, Department of Physics, University of Cambridge, JJ Thomson Ave, Cambridge, United Kingdom}%

\author{Ke Wang}
\affiliation{Cavendish Laboratory, Department of Physics, University of Cambridge, JJ Thomson Ave, Cambridge, United Kingdom}%

\author{Min Lin}
\affiliation{Cavendish Laboratory, Department of Physics, University of Cambridge, JJ Thomson Ave, Cambridge, United Kingdom}%

\author{Boyao Liu}
\affiliation{Cavendish Laboratory, Department of Physics, University of Cambridge, JJ Thomson Ave, Cambridge, United Kingdom}%

\author{Yiru Zhu}
\affiliation{Department of Materials Science and Metallurgy, University of Cambridge, 27 Charles Babbage Road, Cambridge, United Kingdom}%

\author{Ismail Sami}
\affiliation{Department of Materials Science and Metallurgy, University of Cambridge, 27 Charles Babbage Road, Cambridge, United Kingdom}%

\author{Kenji Watanabe}
\affiliation{Research Center for Electronic and Optical Materials, National Institute for Materials Science, 1-1 Namiki, Tsukuba 305-0044, Japan}

\author{Takashi Taniguchi}
\affiliation{Research Center for Materials Nanoarchitectonics, National Institute for Materials Science, 1-1 Namiki, Tsukuba 305-0044, Japan
}

\author{David Ward}
\affiliation{Cavendish Laboratory, Department of Physics, University of Cambridge, JJ Thomson Ave, Cambridge, United Kingdom}%

\author{Andrew Jardine}
\affiliation{Cavendish Laboratory, Department of Physics, University of Cambridge, JJ Thomson Ave, Cambridge, United Kingdom}%

\author{Akshay Rao}
\affiliation{Cavendish Laboratory, Department of Physics, University of Cambridge, JJ Thomson Ave, Cambridge, United Kingdom}%

\author{Manish Chhowalla}
\affiliation{Department of Materials Science and Metallurgy, University of Cambridge, 27 Charles Babbage Road, Cambridge, United Kingdom}%

\author{Sam Lambrick}
\affiliation{Cavendish Laboratory, Department of Physics, University of Cambridge, JJ Thomson Ave, Cambridge, United Kingdom}%
\affiliation{ISIS Facility, Rutherford Appleton Laboratory, Chilton, Didcot, Oxfordshire OX11 0QX, United Kingdom}

\date{\today}

\begin{abstract}
\noindent Two-dimensional (2D) materials have attracted wide-spread interest due to their unique and tunable properties. Their optoelectronic, mechanical, and thermal properties are greatly influenced by crystal defects, which are, in turn, used to control these properties. However, experimental quantification of the density of defects, whether deliberately introduced or inherent, is very difficult in these atomically thin materials. Here we show that helium atom micro-diffraction can be used to measure the defect density in $\sim15\times\SI{20}{\micro\metre}$ monolayer MoS\textsubscript{2}, a prototypical 2D semiconductor, quickly and easily compared to standard methods. We present a simple analytic model, the lattice gas equation, that captures the relationship between atomic Bragg diffraction intensity and defect density. The model, combined with \emph{ab initio} scattering calculations, shows that our technique can immediately be applied to a wide range of 2D materials, independent of sample chemistry or structure. Additionally, wafer-scale characterization is immediately possible.
\end{abstract}

\maketitle


Control of defect density in semiconductors is essential for both current and future optoelectronic devices. In particular, the properties of two-dimensional semiconductors, such as the prototypical MoS\textsubscript{2}, can be tuned using single-atom defects \cite{Regan2022EmergingHeterobilayers} for catalysis \cite{Yang2019SingleCatalysis} and optoelectronic applications \cite{Mitterreiter2021TheMoS2,Barthelmi2020AtomisticMoS2,Chakraborty2019AdvancesMaterials}. The density of defects is often critical in device design, where precise control of the defect density is necessary to produce the desired device properties reproducibly \cite{Zhu2023Room-TemperatureDisulfide}. However, quantification of defect densities in 2D materials remains a significant experimental challenge, with typically used methods (see Table S1) being low temperature photoluminescence (PL), beamline XPS \cite{Zhu2023Room-TemperatureDisulfide} and STEM \cite{Yang2019SingleCatalysis}, with conductive AFM (cAFM) \cite{Xu2023ValidatingMaterials} being explored recently. However, all these methods commonly require complicated sample preparation processes, long measurement or beam line access times and can be prohibitively expensive. As such, there is a characterization shortcoming that will only grow as devices using defect-tuned 2D materials \cite{Lopez-Sanchez2013UltrasensitiveMoS2} gain further traction in both academic research and industry.
Defect characterization over wafer-scale monolayers, which are crucial for commercial and industrial uptake of 2D materials, is particularly difficult with existing techniques. 


Thermal energy, neutral helium atoms provide a uniquely surface sensitive probe that has an interaction cross-section that is fundamentally independent from a sample’s bulk. The helium scatters from the valence electron density of the surface with a classical turning point $\sim2-\SI{3}{\angstrom}$ above the ionic cores \cite{Holst2021}. This means that the scattered helium signal is independent of sample thickness, unlike alternative techniques that use photon or electron probes.  Thus, the helium signal is solely a function of the electronic order at the surface with no beam penetration to sub-surface layers, making it ideal for the characterization of 2D, soft and electrically sensitive materials, regardless of their bandgap, without sample damage or need for specific sample preparation \cite{BarrImagingMicroscopy,Koch2008ImagingMicroscope,T.A.Myles3DMicroscope,Lambrick2020MultipleMicroscopy,Bhardwaj2023ContrastBeams,Radic20243DAtoms,Radic2025HeliometricProfilometry}. Additionally, thermal energy helium beams are highly sensitive to atomic-scale features due to their commensurate de Broglie wavelength ($\sim\SI{0.06}{\nano\metre}$ at $\SI{64}{\milli\electronvolt}$) and the strong attractive component of the helium-surface interaction potential \cite{Poelsema1989ScatteringSurfaces}. The cross sections to monatomic/diatomic adsorbates are therefore many times larger than a typical unit cell, ranging from $30-\SI{200}{\angstrom\squared}$ \cite{Farias1998AtomicSurfaces}. One can expect, therefore, that the scattering cross-section between helium and an atomic vacancy would be similarly enhanced.
Recent work has shown that existing scanning helium microscopes (SHeM) \cite{BarrImagingMicroscopy,Koch2008ImagingMicroscope,Witham2011AMicroscopy} can perform Bragg diffraction to determine the crystallographic structure of a surface with few-micron spatial resolution \cite{vonJeinsen20232DSpot,Hatchwell2024MeasuringMicroscopy}. This technique is termed helium atom micro-diffraction (HAMD).

In this work, we employ HAMD to measure the vacancy-type defect density in mechanically exfoliated monolayer MoS\textsubscript{2} ($\sim15\times\SI{20}{\micro\metre}$) with zero damage or specific sample preparation, and capture the results with a simple model. We also present \emph{ab initio} calculations that, together with the model, validate that our method can immediately be applied to a wide range of materials, independent of sample chemistry or structure. Our method also enables wafer-scale measurements of defect density, where micron scale resolution is not necessary, due to favourable scaling of intensity with increasing sample size\cite{BerginAMicroscope}. 



\begin{figure}[h]
    \centering
    \includegraphics[width=86mm]{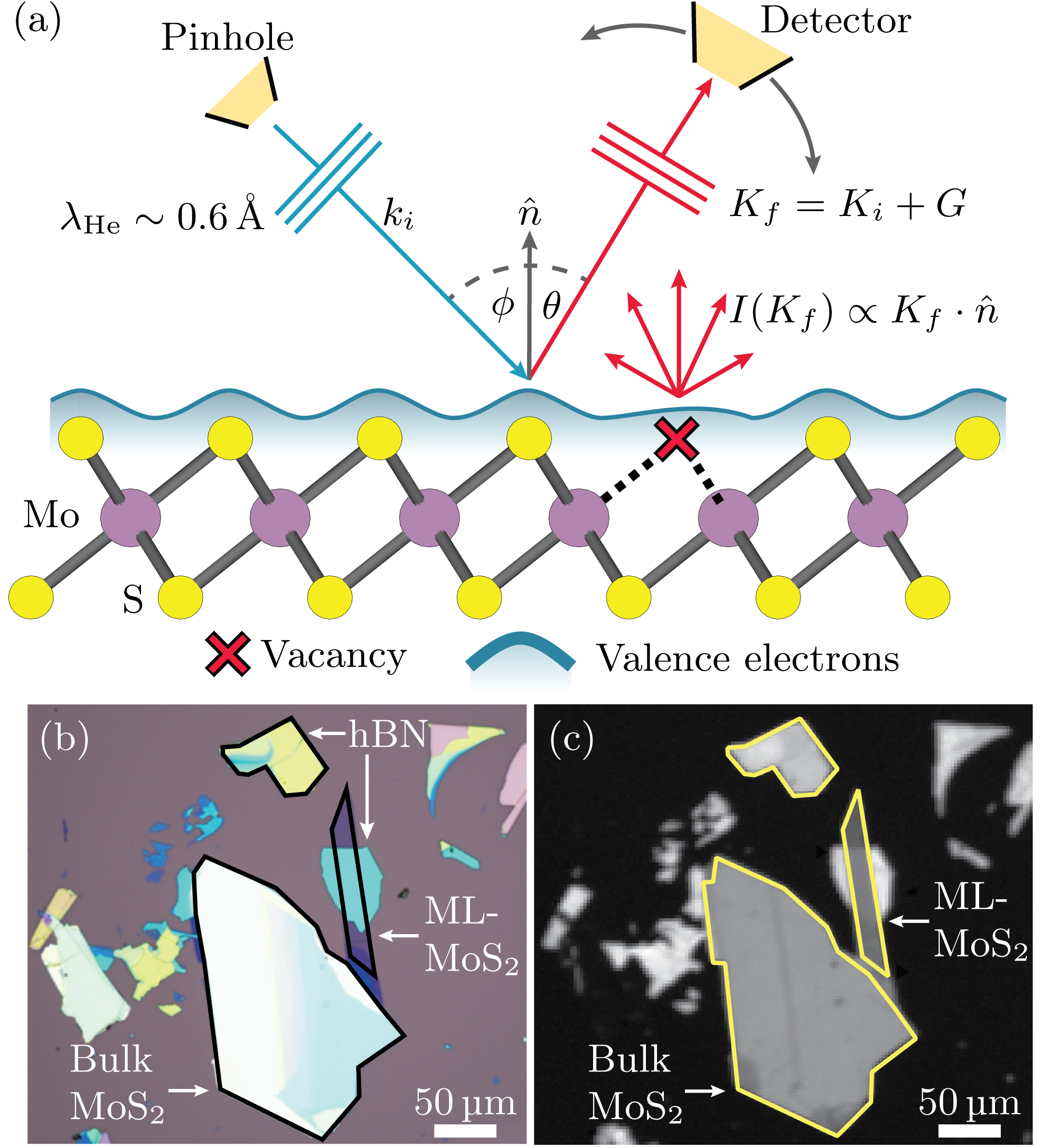}
    \caption{(a) A matter wave scattering from the outermost electron density of a 2D material. Most of the scattered flux is directed into kinematic channels (ordered, Bragg diffraction). Defects in the surface introduce disorder, resulting in some flux being scattered diffusely, following a cosine-like distribution. The ratio of ordered to disordered scattering therefore encodes the degree of order at the surface. The instrument used has a fixed angle of incidence ($\phi=\SI{45}{\degree}$) and varies the outgoing detection angle ($\theta$) by sample manipulation. An optical image (b) and helium micrograph (c) show the typical sample layout.}
    \label{fig:1}
\end{figure}

A helium matter wave can diffract from the surface corrugation created by the valence electron density, or long-range electronic order, of the solid, as shown in Figure \ref{fig:1}, resulting in Bragg diffraction that encodes atomic scale information on the surface structure \cite{Farias1998AtomicSurfaces}. Measurement of long-range electronic order is the ultimate determining factor in a material’s optical and electronic properties, allowing helium scattering to directly probe that which determines device performance. Historically, such helium diffraction measurements were limited to millimetre-scale, carefully prepared, single crystal samples; however, recent advances in instrumentation \cite{vonJeinsen20232DSpot} have allowed microscopic atom diffraction for the first time. A diffraction pattern is measured by scanning the scattered helium flux (with a high-sensitivity custom atom detector \cite{Bergin2021Low-energySpecies}) as a function of the in-plane momentum transfer, $\Delta K$($\theta$).  Our instrument uses manipulations of the sample position and orientation to allow scanning through the outgoing beam angle, $\theta$, and hence $\Delta K$. 

The instrument can also acquire images in which the scattering distribution is frozen at a single $\Delta K$ and allows us to acquire diffraction-optimised micrographs \cite{vonJeinsen20232DSpot,Bergin2020ObservationMicroscopy}. Micrographs allow for qualitative discrimination between different materials and crystalline orientations. A micrograph of a sample used in this work is show in Figure \ref{fig:1}(c). Full details on the sample are contained in the Supplemental Materials.

A schematic of a sulfur vacancy in MoS\textsubscript{2} is shown in Figure \ref{fig:1}(a). The defect disrupts surface periodicity and causes a reduction in the intensity of helium flux that is scattered into diffraction channels, instead scattering the signal into a cosine-like background\cite{Lambrick2022ObservationMicroscopy,Staszewski2023BlueScattering}. The method is therefore a purely geometric quantification of the degree of order at a surface making, it agnostic to a sample’s chemical, optical and electrical properties. 

An increasing fractional defect density, $\Theta$, will therefore result in a decreasing intensity within diffraction peaks. $\Theta=\nicefrac{n_V}{n_S}$ where $n_V$ and $n_S$ are the actual areal densities of defects and sulfur sites on the top surface ($n_S=\SI{1.16e15}{\per\centi\metre\squared}$), respectively. The relationship is captured by the lattice gas equation,
\begin{equation}
    \frac{I}{I_0}=(1-\Theta)^{\nicefrac{2S}{A_{\mathrm{cell}}}}
\label{eqn:lattice_gas}
\end{equation}
where $\Theta$ is the defect density expressed as a fraction of the available sites on the surface, $S$ is the `quantum' or `amplitude' cross section and $A_\mathrm{cell}$ is the unit cell area \cite{Poelsema1989ScatteringSurfaces}. It is important to notice that, for a given fractional defect density ($\Theta$) the differential intensity is solely dependent on the ratio of the He-defect scattering cross-section ($S$) to the primitive unit cell area $\nicefrac{S}{A_{\mathrm{cell}}}$ ($A_{\mathrm{cell}}=\SI{8.6}{\angstrom\squared}$ for MoS\textsubscript{2}). The cross-section $S$ is, for helium scattering, not strictly correlated to the physical size of the defect due to long-range He-defect interactions being much larger than the hard-shell radius\cite{Poelsema1989ScatteringSurfaces}, an affect termed the `giant cross-section'. The giant cross-section is unique to low energy neutral atoms and is key in achieving high sensitivity to low defect densities. The higher-order polynomial dependence of equation (\ref{eqn:lattice_gas}) leads to increasing sensitivity to defects as $\Theta\rightarrow0$. The method can be applied, without adjustment to theory or experiment, to any other 2D material with vacancy-type defects. Although assuming flux scatters from a disordered surface according to a cosine-like distribution centred on the surface normal is a good approximation \cite{Lambrick2022ObservationMicroscopy}, in principle the exact scattering distribution depends on the type of defect/adsorbate \cite{Poelsema1989ScatteringSurfaces}. Therefore it may be practicable in the future to distinguish between defect types using HAMD – including multi-vacancies, 1D and 2D defects. 

By reference to the intensity from a pristine sample, $I_0$, the defect density can be inferred from diffraction intensity, $I$. We note that while equation (\ref{eqn:lattice_gas}) is valid for cross-sections larger than the unit cell ($S\gg A_\mathrm{cell}$), it does assume isolated, non-interacting defects and may break down for very high defect densities where nearby He-defect scattering cross-sections overlap significantly. Later in this Letter, we further discuss the isolated-defect assumption and examine its validity by comparison with a model that explicitly incorporates overlap between defect scattering cross-sections.

Our test of the lattice gas equation is therefore particularly stringent, as the highest defect density measured approaches the upper limit of what is generally considered compatible with high-performance electronic performance in TMDs\cite{Wang2024CriticalDichalcogenides,HongExploringMonolayers}. In addition, TMDs exhibit relatively high defect densities ($\SI{e12}{}$–$\SI{5e14}{\per\centi\metre\squared}$) compared to other two-dimensional materials \cite{Zhu2023Room-TemperatureDisulfide,Zhang2024Low-Defect-DensityStrategy}. Materials such as graphene or hBN commonly display densities in the range $\SI{e9}{}$–$\SI{e13}{\per\centi\metre\squared}$, depending on fabrication method \cite{Edelberg2019ApproachingControl,Gong2023CoherentNitride}. Consequently, the method's successful application to TMDs suggests that the approach is readily extendable to other 2D systems where defect densities are lower and lattice-gas assumptions hold.

We present complementary calculations in the End Matter based on a DFT-fitted He--surface interaction potential (Fig.~\ref{fig:defect_PES_equipotentials}), using close-coupling elastic scattering calculations\cite{ManolopoulosIterativeApproach,Riley2008AnalysisResonances,Avidor2016HeliumSurfaces,Staszewski2023BlueScattering} to determine the diffraction intensities associated with pristine and defective MoS\textsubscript{2} surfaces. These calculations show that increasing disorder redistributes scattered intensity away from the ordered Bragg peaks and into diffuse scattering, and are not used to infer experimental defect densities. The calculated cross sections are of similar magnitude to the experimental values, although on average they are \(\sim25\%\) smaller. The close-coupling treatment assumes a frozen surface; however, thermal attenuation changes only the absolute calculated intensities, while the cross-section is extracted from normalised experimental intensities $(I/I_0)$ and is unchanged within uncertainty at $\SI{120}{\degreeCelsius}$ and $\SI{200}{\degreeCelsius}$ (Fig. S3).

Three ML-MoS\textsubscript{2} samples of increasing defect density were prepared using thermal annealing under an argon/hydrogen (95\%/5\%) atmosphere to induce sulfur vacancies. We replicate the annealing procedures by Zhu \emph{et al.} \cite{Zhu2023Room-TemperatureDisulfide} precisely to ensure reliable defect densities. The monolayer flakes were then placed on a $\sim\SI{25}{\nano\metre}$ thick hBN buffer. The hBN protects the morphological and electronic properties of the ML-MoS\textsubscript{2}, as previously shown using LEEM/D and PL \cite{Man2016ProtectingBuffer}. The MoS\textsubscript{2}-on-hBN sample is mounted on a SiO\textsubscript{2}. Full sample details are discussed in Supplemental Materials.

\begin{figure}[h]
    \centering
    \includegraphics[width=\linewidth]{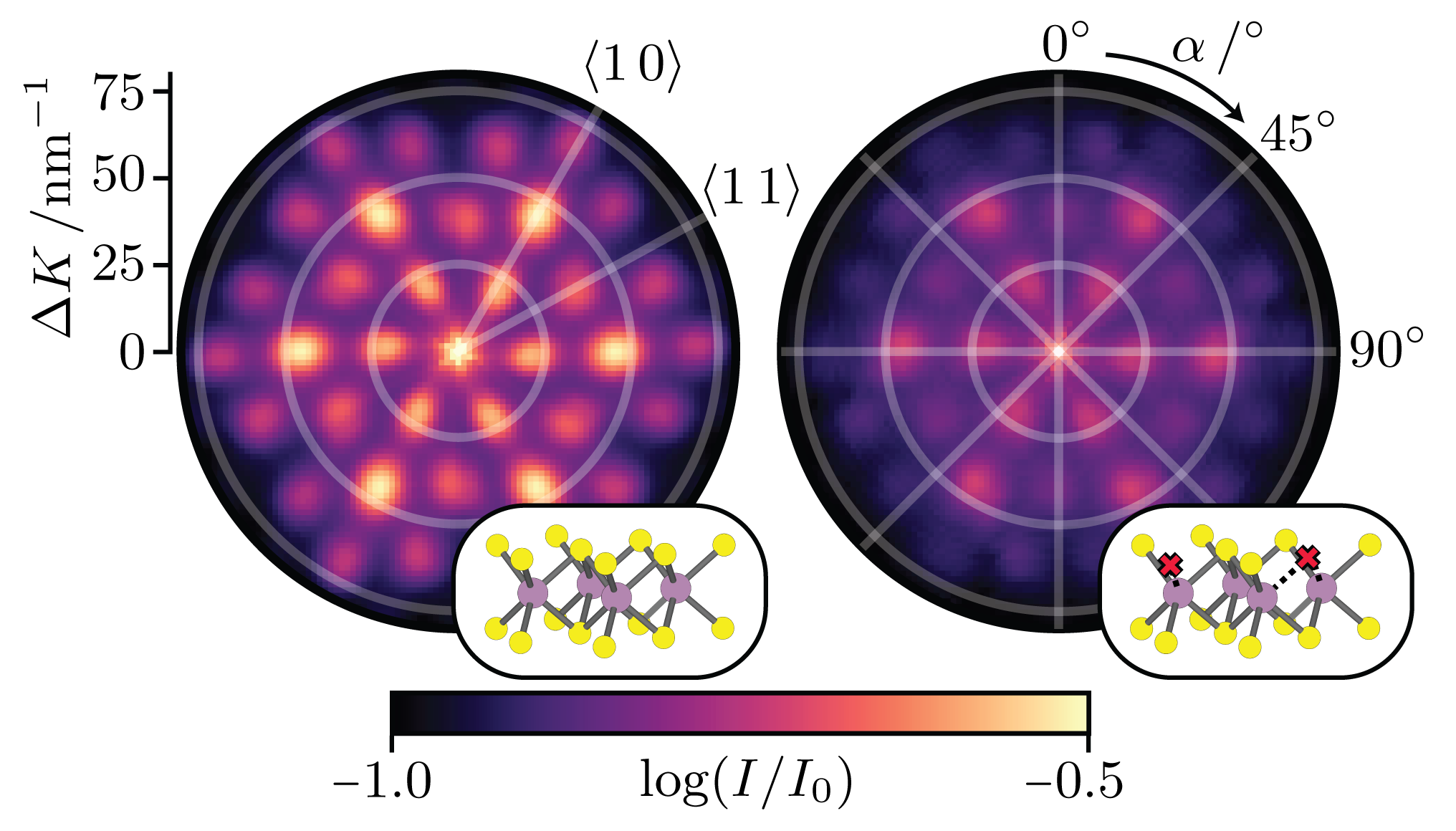}
    \caption{2D diffraction scan of lowest (left) and highest (right) defect densities of monolayer MoS\textsubscript{2} measured in this study. The native defect density (left, $n_V=\SI{0.1e14}{\per\centi\metre\squared},\,\Theta_1=0.01$) produces more intense diffraction peaks at all orders in comparison to the high defect density (right, $n_V=\SI{1.8e14}{\per\centi\metre\squared},\,\Theta_3=0.16$) sample. Schematics of pristine and defective ML-MoS\textsubscript{2} surface are inset in the corresponding diffraction pattern.}
    \label{fig:2}
\end{figure}


2D micro-diffraction patterns for the least ($\Theta_1=0.01$) and most ($\Theta_3=0.16$) defective samples are shown in Figure \ref{fig:2}. All diffraction measurements presented were acquired at a sample temperature of $\SI{120}{\degreeCelsius}$ after initial \emph{in situ} annealing at $\SI{220}{\degreeCelsius}$ for 2 hours to clean the sample surface. Since MoS\textsubscript{2} is stable under inert/vacuum atmosphere up to $\sim\SI{750}{\kelvin}$ \cite{Loi2020GrowthStudy}, there is no risk of heating-induced change to the samples. Diffraction data is acquired at an elevated temperature to ensure the cleanliness of the sample surface from adsorbates. We demonstrate experimentally that measurement of defect density is independent of sample temperature by fitting the lattice gas equation (Eqn. \ref{eqn:lattice_gas}) to diffraction scans taken at $\SI{120}{\degreeCelsius}$ and $\SI{200}{\degreeCelsius}$ in figure S3 (Supplementary Materials). Figure \ref{fig:2} shows the parallel momentum transfer (labelled $\Delta K$) radially, with the orientation of the crystal surface (labelled $\alpha$) azimuthally, the patterns can be considered a representation of the reciprocal surface lattice. The trigonal sulfur surface lattice is clearly seen, and we find the sulfur-sulfur spacing on the surface of the ML-MoS\textsubscript{2} as $3.20\pm\SI{0.07}{\angstrom}$ compared to $\sim\SI{3.15}{\angstrom}$ in literature \cite{Nguyen2019FabricationLIBs}. Details of lattice constant measurement are shown in Figure S2. Figure \ref{fig:2} shows a clear decrease in flux scattered into kinematic channels between the two patterns as defect density is increased, while their relative positions remain unchanged.

\begin{figure}[h]
    \centering
    \includegraphics[width=\linewidth]{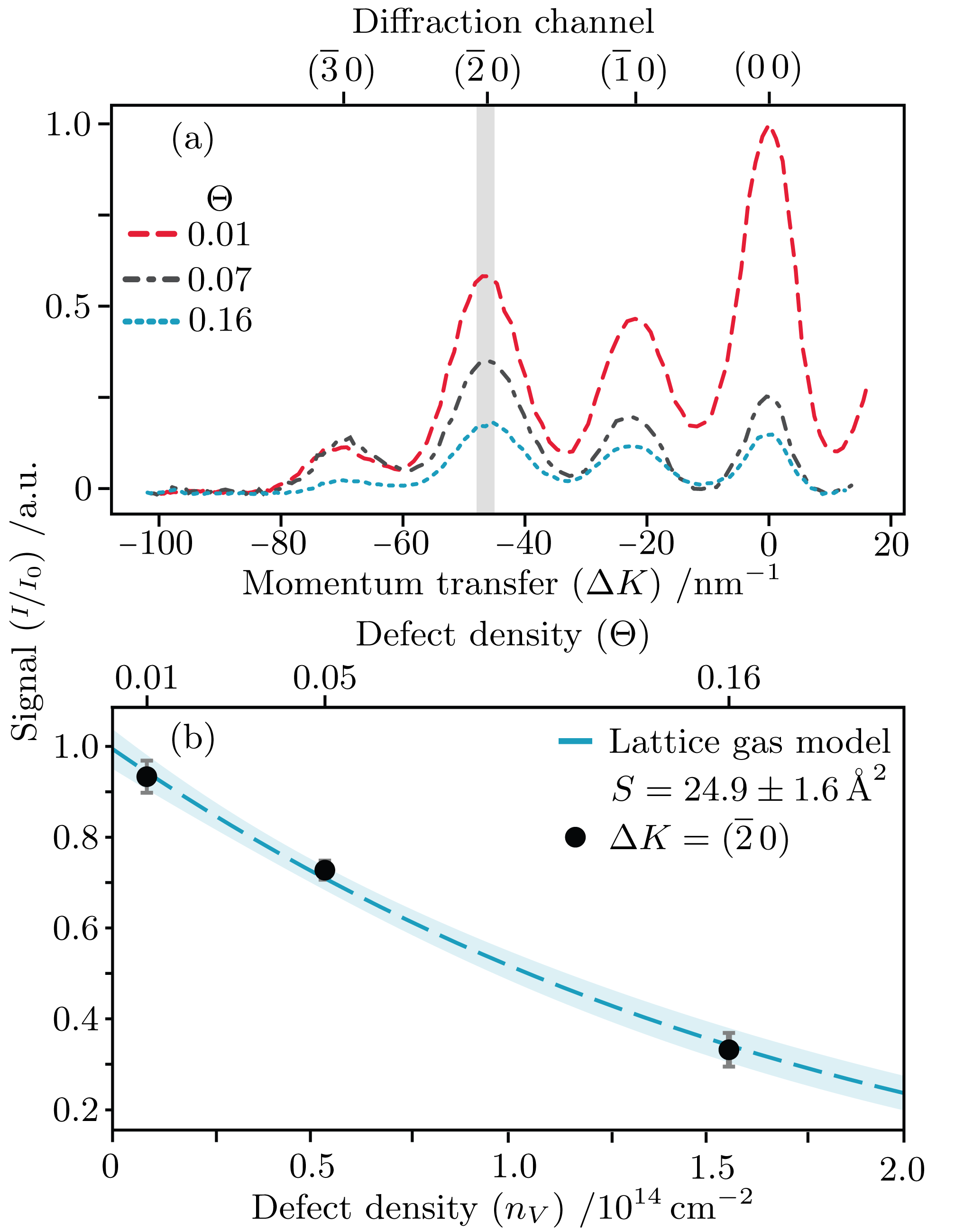}
    \caption{(a) Diffraction scans along \hkl<10> of increasing defect density ML-MoS\textsubscript{2} acquired at $\SI{120}{\degreeCelsius}$. (b) Fitting the lattice gas equation to the intensities of the \hkl(-20) peaks yields a He-defect scattering cross-section $S_{LG}=24.9\pm\SI{1.6}{\angstrom\squared}$, approximately $\times2.9$ larger than the unit cell area, $A_{cell}$.}
    \label{fig:3}
\end{figure}

The electronic order of the surface is, broadly speaking, sensitive to three parameters that describe a potential energy surface (PES). The electronic corrugation (i) and lattice constant (ii) are primarily responsible for changing relative diffraction peak intensities and their positions in $\Delta K$, respectively; they do not change as the defect density is varied in the sample. As such, changes in absolute diffracted intensity, as seen in Figure \ref{fig:2} are attributed to changing defect density ($\Theta$) (iii) and can therefore be modelled using equation (\ref{eqn:lattice_gas}). Further discussion on the PES is presented in the End Matter. 
Figure \ref{fig:3}a presents 1-dimensional diffraction scans were taken along the principal azimuth (\hkl<10> direction shown in Figure \ref{fig:2}) for each sample. The measurements were normalised using a diffusely scattered background measured on the SiO\textsubscript{2} substrate to account for fluctuations in detection sensitivity. Helium scattering from SiO\textsubscript{2} (purple/black backgrounds in Figure \ref{fig:1}(b) and (c), respectively) is almost entirely disordered (cosine-like scattering distribution) and thus produces significantly lower intensity signals than MoS\textsubscript{2} or hBN which are highly reflective. We extract the intensity of the diffracted helium intensity by Gaussian fitting of the \hkl(-10) and \hkl(-20) peaks. The relationship of intensity to defect density of \hkl(-20) is shown in Figure \ref{fig:3}(b), and the \hkl(-10) equivalent shown in Supplemental Materials (Figure S4). The \hkl(00) and \hkl(-30) peaks are not used here due to their high sensitivity to sample tilt and low intensity, respectively, although in principle the method works for any diffraction condition, including those with momentum transfers beyond the scattering plane.

To test the validity of the lattice gas model (equation \ref{eqn:lattice_gas}) we compare to a model that accounts for `medium overlap' between defect cross-sections on a hexagonal lattice (equation S1) derived by Poelsema and Comsa\cite{Poelsema1989ScatteringSurfaces}.
Fitting both models to the \hkl(-20) peak yields empirical helium-defect cross-sections of $S_{LG}=24.9\pm\SI{1.6}{\angstrom\squared}$ and $S_{MO}=25.3\pm\SI{1.7}{\angstrom\squared}$, comparable to, but somewhat smaller than monatomic adsorbates\cite{Farias1998AtomicSurfaces}. For the \hkl(-10) peak the models return cross-sections $S_{LG}=29.8\pm\SI{4.5}{\angstrom\squared}$ and $S_{MO}=30.5\pm\SI{4.2}{\angstrom\squared}$. It is notable that the cross-sections are significantly larger than the area of a unit cell ($\SI{8.6}{\angstrom\squared}$ for MoS\textsubscript{2}), highlighting the sensitivity of helium atom scattering. Cross section values are not necessarily expected to be the same for every diffraction peak due to the long range attractive potential and different atom trajectories through it\cite{Poelsema1989ScatteringSurfaces}. The fit of both models and the cross-sections they report are in close agreement for both the \hkl(-10) and \hkl(-20) peaks (Figure S1, End Matter). We therefore conclude that the lattice gas model is a good approximation for highly defective TMD systems, and recommend its use given we study the upper range of defect density of technological interest.
While there is no single upper threshold for defect density in TMDs, device studies consistently indicate that densities beyond the mid-$\SI{e13}{\per\centi\metre\squared}$ ($\Theta\sim0.04$) range severely degrade key electronic properties, such as carrier mobility, rendering the material unsuitable for high-performance optoelectronic devices \cite{Gali2020ElectronicVacancies,Chen2018TuningEngineering,Wang2024CriticalDichalcogenides}. While morphological order is substantially more robust than electronic order in TMDs, experimental studies indicate that lattice coherence breaks down at defect densities $\sim\SI{5e14}{\per\centi\metre\squared}$ ($\Theta\sim0.43$), with MoS\textsubscript{2} undergoing a `complete crystalline-to-amorphous transition'\cite{Burns2020StrainS2}.


Fitting to the lattice gas model, equation (\ref{eqn:lattice_gas}), shown in Figure \ref{fig:3}b acts as a calibration curve for all ML-MoS\textsubscript{2} because the model is solely dependent on the He-defect scattering cross-section, $S$. One can look towards high-throughput quality control of 2D semiconductors by integrating HAMD into molecular beam epitaxy (MBE) or chemical vapour deposition (CVD) systems directly as an \emph{in situ} and \emph{operando} measurement of defect density immediately during wafer growth. Owing to its non-destructive, chemically inert beam and ability to measure intrinsic, unprepared semiconductors, the wafers would be unaffected by the measurement and can go on to be used in further manufacturing steps. The same apparatus could also simultaneously extract key parameters on the electronic and vibrational structure of the material, such as electron-phonon coupling constants and thermal expansion coefficients \cite{Manson2022AtomSurfaces,Anemone2019ElectronPhononScattering}. The key innovations, in instrumentation, that enable the few-micron spot size in the current work are recent advances in the generation of intense neutral atom sources \cite{BerginAMicroscope,Kelsall2024MinimisingSources} and custom high-sensitivity atom detectors \cite{Bergin2021Low-energySpecies}. 
For future work where quality control of wafer-scale samples may be desired the micron-scale spatial resolution demonstrated in the current work, which uses small exfoliated flakes common in research environments, may not be required. For larger samples the aggressive scaling of intensity, $I$, with spot size, $r$, in SHeM, of $I\propto r^4$ \cite{BerginAMicroscope}, may prove to be favourable as significant increases in signal will be achieved with minor sacrifices to spatial resolution (spot size). With the resulting increase in signal, detection can be performed using inexpensive, generic mass spectrometers, while simultaneously reducing measurement times and improving accuracy, both of which are ultimately determined by the achievable signal-to-noise ratio.


We have demonstrated helium atom micro-diffraction (HAMD) as a direct and non-destructive method for the measurement of vacancy-type defect density in device-scale ($\sim15\times\SI{20}{\micro\metre}$) monolayer TMDs using MoS\textsubscript{2}. These results position HAMD as an indispensable tool for the characterization of defect density in 2D materials spanning the microscopic and wafer scales to support the ever-growing academic and industrial research and development of 2D material-based devices.

We present an analytic model, the lattice gas equation, that accurately captures the relationship between diffraction intensity and vacancy-type defect density in a monolayer semiconductor. We also performed \emph{ab initio} calculations that show, together with the lattice gas model, that the method’s contrast is independent of surface chemistry or structure. This allows it to be applied immediately to a wide range of systems whose optoelectronic, mechanical or thermal properties can be modulated via surface defects or dopants, examples include hBN \cite{Stern2024AConditions}, graphene \cite{Bhatt2022VariousReview}, doped diamond \cite{Einaga2022Boron-DopedApplications}, alongside other TMDs. 

Leveraging recent advances in neutral atom beam generation and detection, we demonstrate the method under its most demanding conditions by achieving micron-scale spatial resolution ($\SI{5}{\micro\metre}$ beam spot size). Owing to the favourable scaling of detected intensity with beam spot size, HAMD is scalable and readily compatible with wafer-scale growth techniques, enabling \emph{in situ} and \emph{operando} quality control essential for the widespread adoption of 2D materials in electronic devices.

Future work could leverage the technique’s demonstrated sensitivity to surface electronic structure to investigate a range properties that also have significant effects on device performance and are encoded within atomic scattering, such as surface contamination, thermal expansion coefficients and electron-phonon coupling in monolayer materials \cite{Manson2022AtomSurfaces,Anemone2019ElectronPhononScattering,Anemone2018ExperimentalMoS2,Mak2010AtomicallySemiconductor}, with accessible characterization areas spanning microns to centimetres. 

\section*{Data availability}
The data underlying all figures in the main text and Supplementary Information is available at \url{https://doi.org/10.5281/zenodo.17551164}. All code used in this work is available from the corresponding authors upon reasonable request.

\begin{acknowledgments}
The work was supported by EPSRC grant EP/R008272/1, Innovate UK/Ionoptika Ltd. through Knowledge Transfer Partnership 10000925. The work was performed in part at CORDE, the Collaborative R\&D Environment established to provide access to physics related facilities at the Cavendish Laboratory, University of Cambridge and EPSRC award EP/T00634X/1. S.M.L. acknowledges support from EPSRC grant EP/X525686/1. K. Watanabe and T.T. acknowledge support from the JSPS KAKENHI (Grant Numbers 21H05233 and 23H02052), the CREST (JPMJCR24A5), JST and World Premier International Research Center Initiative (WPI), MEXT, Japan. K. Wang acknowledges the Cambridge Trust and the CSC for financial support. The authors thank Christoph Schnedermann for useful discussions.
\end{acknowledgments}

\clearpage
\appendix

\renewcommand{\thefigure}{A\arabic{figure}}
\setcounter{figure}{0}

\renewcommand{\thetable}{A\Roman{table}}
\setcounter{table}{0}

We evaluate theoretical He-defect cross-sections from close-coupled (CC) scattering calculation diffraction intensities in MoS\textsubscript{2}. The calculations use an analytic He–surface interaction potential fitted to a sparse grid of DFT energies, with a finite-temperature correction applied to the resulting intensities.

\vspace*{-1\baselineskip}

\section{He-MoS2 interaction potential calculation}
\label{sec:construct_potential}
DFT calculations were performed at the high-symmetry hollow (\(H\)), sulfur (\(S\)), and molybdenum (\(M\)) sites on the surface. For \(i=H,S,M\), the interaction energy between a single helium atom and the MoS\textsubscript{2} surface was evaluated as a function of helium height \(z\), yielding \(V_i(z)\) within the rigid-surface approximation, in which the helium atom is assumed not to significantly perturb the substrate. We used a \(4\times4\times1\) supercell to avoid spurious interactions between adjacent He atoms. All calculations used CASTEP \cite{Clark2005FirstCASTEP} with the PBE functional \cite{Perdew1996GeneralizedSimple} and the DFT-D3 \cite{Moellmann2014DFT-D3Crystals} dispersion correction; further details are given in the Supplementary Materials.

\begin{figure}
    \vspace{\baselineskip}
    \includegraphics[width=86mm]{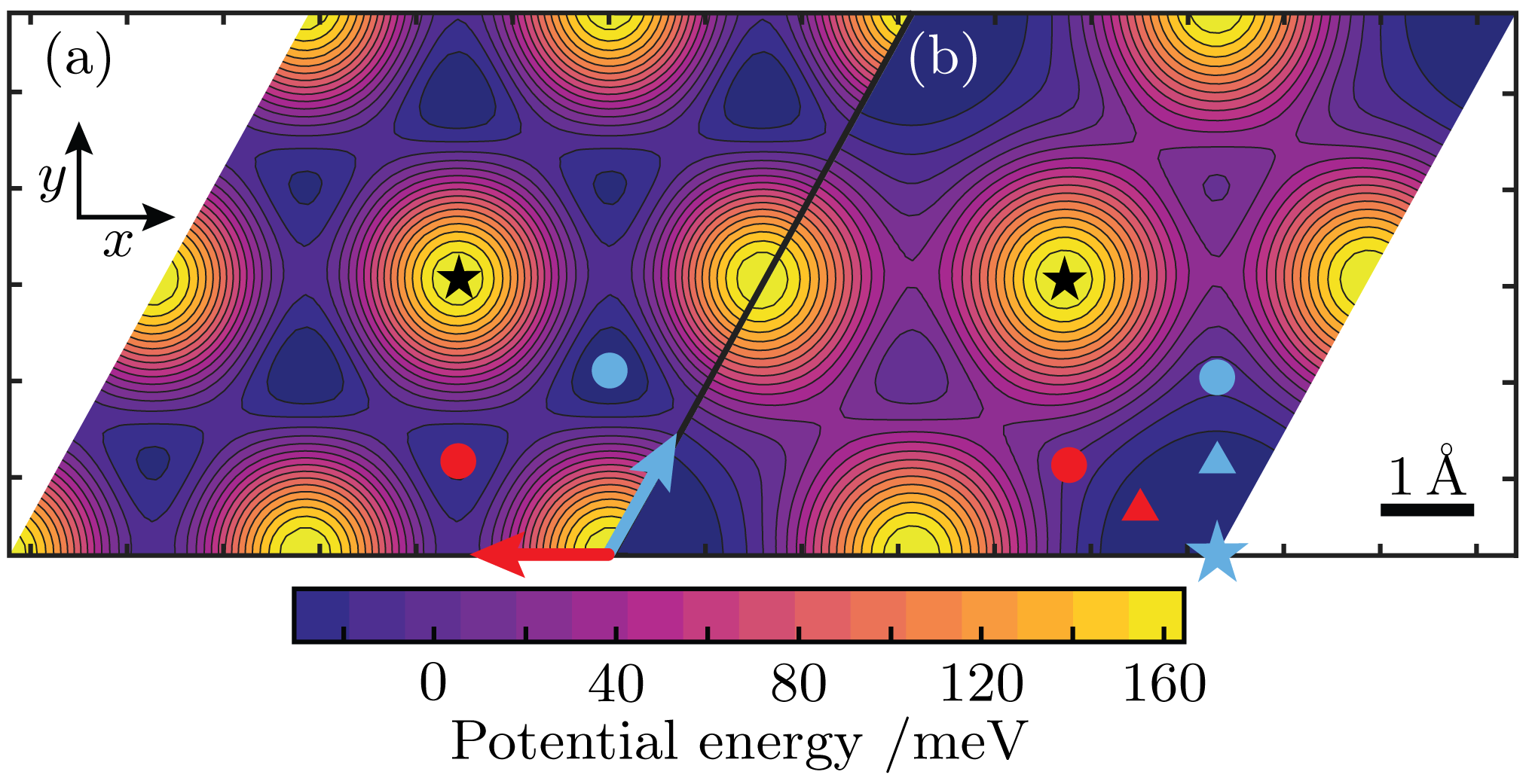}
    \caption[Defect-free (a) and defective (b) He-MoS\textsubscript{2} potential energy surfaces]{He-MoS\textsubscript{2} potential energy surfaces at a height $z=\SI{2.5}{\angstrom}$ from the top sulfur ionic cores, approximately the classic turning point of the incident helium atom. In (a) the blue dot marks a hollow site and the red dot marks a molybdenum atom site. In (b) The blue triangle marks a bridge site between the defect site and the empty site and the red triangle marks a bridge site between the defect site and the molybdenum atom site. The arrows in (a) show the real-space lattice vectors. In (b) a sulfur vacancy is marked with a blue star. The black star represents a sulfur site.}
    \label{fig:defect_PES_equipotentials}
\end{figure}

 For a defect-free $\Theta=0$ surface, we use a corrugated Morse-like potential,
\begin{equation}
\begin{aligned}
    V(x,y,z) = &V_S(z) Q(x,y) + V_H(z) Q\!\left(x, y - \frac{c}{\sqrt{3}}\right) \\
          &+ V_M(z) Q\!\left(x - \frac{c}{2}, \; y - \frac{c}{2\sqrt{3}}\right)
\end{aligned}
\label{eq:defect_free_pes}
\end{equation}
where $c$ is the lattice parameter of MoS\textsubscript{2}, $Q$ is the corrugation function and $V_i$ are the modified Morse-like potentials for the three high symmetry sites. The $S$, $M$, and $H$ sites are marked on Figure \ref{fig:defect_PES_equipotentials} which plots a slice through the fitted potential at a height of $z=\SI{2.5}{\angstrom}$ above the sulfur ion cores. The hexagonal corrugation function takes the form,
\begin{equation}\label{eq:potential}
Q(x,y) = \frac{2}{9}[\cos(x' - y')+\cos(2y')+\cos(x' + y' + 3\pi)]
\end{equation}
where $x' = \tfrac{2\pi x}{c}$ and $y' = \tfrac{2\pi y}{c\sqrt{3}}$, such that at specific site $i$ all other terms vanish from the total potential and $V=V_i$. Each modified Morse potential $V_i$  takes the form,
\begin{equation}\label{eq:morse}
V(z) = D_i \Big[
\mathrm{e}^{2\alpha_i (z_{0}-z)}
- 2a_i\, \mathrm{e}^{\alpha_i (z_{0}-z)}
- 2b_i\, \mathrm{e}^{2\beta_i (z_{1}-z)}
\Big],
\end{equation}
where $D_i,a_i,b_i,\alpha_i,\beta_i,z_0,z_1$ are empirical parameters that are determined from fitting to the DFT calculations. The additional attractive exponential term in Eqn. \ref{eq:morse}, parametrized by $b,\,\beta,\,z_1$, improves the description of the attractive tail over the range of z relevant to the scattering calculation compared to a standard Morse potential. The parameters are obtained by fitting to DFT interaction energies calculated as a function of helium height $z$, including DFT-D3 dispersion corrections.
We emphasise that this analytic form is not intended to reproduce the exact long-range asymptotic interaction, instead, providing a smooth interpolation of the DFT-derived potential over the range of distances relevant to the scattering process. For the $\SI{64}{\milli\electronvolt}$ helium beam used here, the classical turning point lies $2-\SI{3}{\angstrom}$ above the top sulfur plane\cite{Holst2021}, meaning that scattering is dominated by the repulsive wall and near-well potentials, while the long-range tail has only a minor influence on the diffraction intensities

For the defective surface, we modify the potential in Eqs.~\ref{eq:potential}--\ref{eq:morse} according to
\begin{equation}
V(\mathbf{r},z)=V_{\mathrm{defect}}(z)\,L(R)+V_{\mathrm{pristine}}(\mathbf{r},z)\,[1-L(R)],
\label{eq:defect_potential}
\end{equation}
where \(R=\sqrt{x^2+y^2}\) is the in-plane radial distance from a defect centre in a supercell. The complete potential used in the calculations is obtained by periodically repeating the supercell. The function \(L(R)\) is a smooth radial logistic switching function defined by
\begin{equation}
f(R)=\frac{1}{1+\exp\!\left(\frac{R-\mu}{\Delta}\right)}, \qquad
L(R)=\frac{f(R)}{f(0)},
\end{equation}
so that \(L(0)=1\) at the defect centre and \(L(R)\rightarrow 0\) at large \(R\). The parameter \(\mu=0.49~\text{\AA}\) sets the characteristic radius of the switching region, while \(\Delta=1/15.9\approx 0.063~\text{\AA}\) controls the sharpness of the transition. This construction localises the defect perturbation while ensuring a smooth interpolation between the defect and pristine potentials, thereby avoiding numerical artefacts associated with a discontinuous potential. This also allows supercells of different size to be constructed using the same local defect potential, such that varying the supercell size corresponds to varying the defect density. At the defect site, the potential \(V_{\mathrm{defect}}(z)\) is used, which takes the form
\begin{equation}\label{eq:pes_defect}
V_{\text{defect}}(z) = d \big[ \mathrm{e}^{2\gamma (z_{2}-z)} 
- 2c\, \mathrm{e}^{\gamma (z_{2}-z)}- 2e\, \mathrm{e}^{2\lambda (z_{3}-z)} \big]
\end{equation}
where $d,c,e,\gamma,\lambda,z_2,z_3$ are parameters fitted to a DFT He-surface potential over the defect site. A slice of pristine and defected $V(x,y,z)$ at $z=\SI{2.5}{\angstrom}$ above the sulfur ion cores is plotted in Fig. \ref{fig:defect_PES_equipotentials}. 
\begin{table}[H]
\begin{ruledtabular}
\begin{tabular}{c c c c c c c c}
Site/$i$ & d/$\SI{}{\milli\electronvolt}$ & $a$ & $b$ & $\alpha\,/\SI{}{\per\angstrom}$ & $\beta\,/\SI{}{\per\angstrom}$ & $z_0\,/\SI{}{\angstrom}$ & $z_1\,/\SI{}{\angstrom}$ \\
\hline
S & 20 & 0.812 & 0.196 & 1.44 & 0.203 & 3.37 & 1.73 \\
M & 20.1 & 1 & 0.0026 & 1.15 & 1.24 & 3.22 & 4.19 \\
H & 25 & 0.464 & 0.199 & 1.1 & 0.648 & 3.14 & 3.82\\\hline
 & $d$ & $c$ & $e$ & $\gamma$ & $\lambda$ & $z_2$ & $z_3$ \\\hline
Defect & 32.8 & 0.0311 & 16.4 & 0.921 & 0.924 & 5.6 & 3.71\\

\end{tabular}
\end{ruledtabular}
\caption{Parameters describing the He-MoS\textsubscript{2} interaction potential fitting to DFT calculations.}
\label{tab:pes_parameters}
\end{table}

\vspace{-1.5\baselineskip}

\section{Scattering calculations}

The He--MoS\textsubscript{2} interaction potentials were used to calculate the elastic diffraction intensities in discrete scattering channels using a close-coupling (CC) method, which solves the time-independent Schr\"{o}dinger equation for helium atom scattering from a periodic surface potential\cite{ManolopoulosIterativeApproach}. We use the \texttt{MultiScat} software package~\cite{CambridgeAtomScatteringCentre2020GitHubCambridge-Atom-Scattering-Centre/Multiscat}. The CC method has previously been applied to helium scattering from ordered periodic surfaces~\cite{Mayrhofer-Reinhartshuber2013Helium-surfaceCalculations,Riley2008AnalysisResonances,vonJeinsen20232DSpot}, while \texttt{MultiScat} has been used explicitly for disordered surfaces~\cite{Staszewski2023BlueScattering} and periodically repeated disordered supercells~\cite{Avidor2016HeliumSurfaces}. The CC method calculates the scattered intensity \(I(\bm{K}_f,\bm{K}_i)\) where \(\bm{K}_i\) and \(\bm{K}_f\) are the incident and final wavevectors projected parallel to the surface. The present calculations employ a frozen-surface potential and therefore do not include finite-temperature effects, although bound-state resonances are retained.

\begin{figure}
    \centering
    \includegraphics[width=86mm]{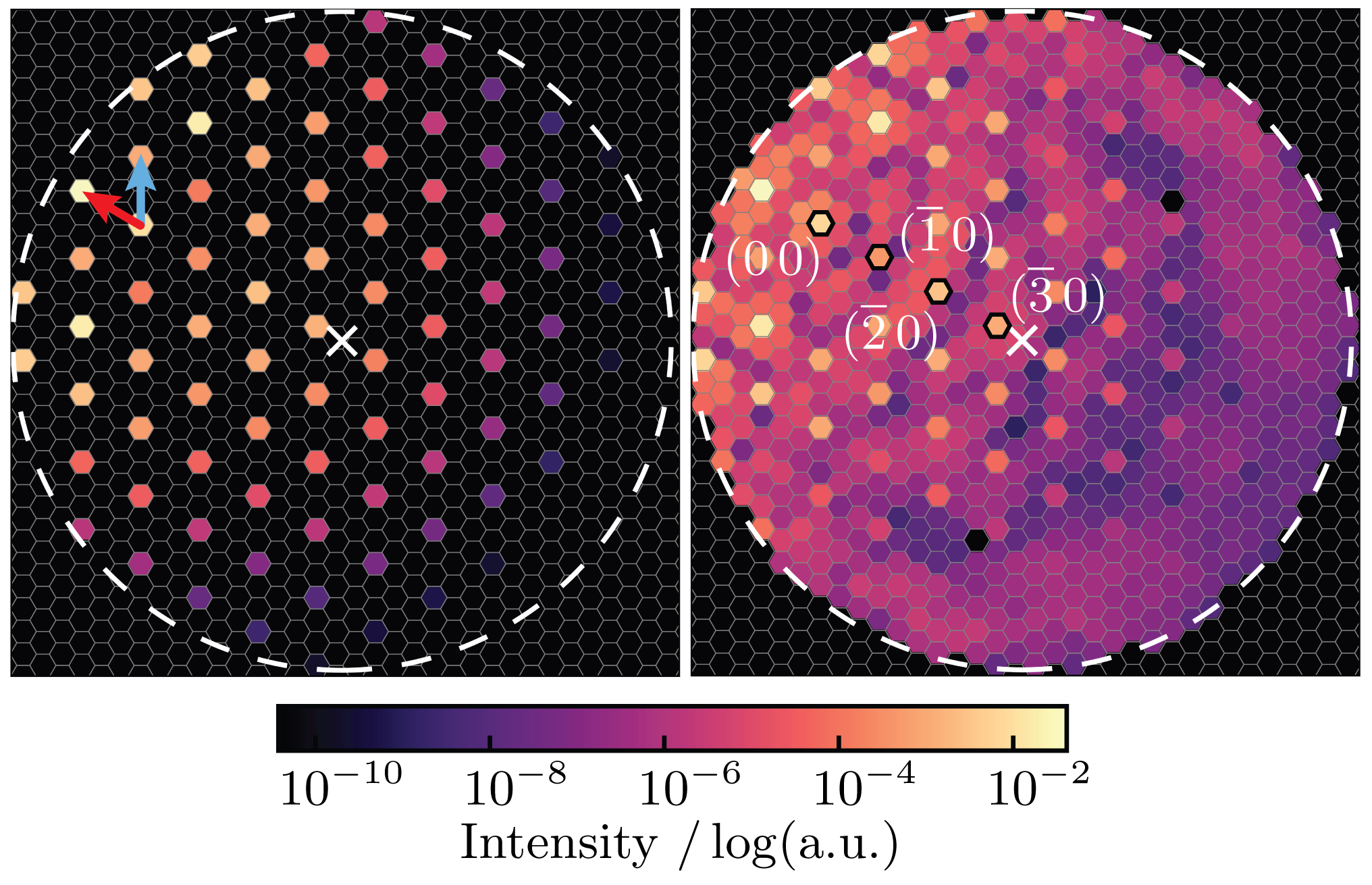}
    \caption{CC-calculated reciprocal-space distribution of scattered intensity for a \(3\times3\) supercell using the DFT-fitted potentials with thermal attenuation applied according to Eq.~\ref{eq:dw}. Left: pristine surface. Right: defective surface. Each hexagon represents the intensity of a diffraction channel centred on its outgoing wavevector. The specular \(\hkl(00)\) condition is off-centre due to the \(\SI{45}{\degree}\) incidence geometry. Channels beyond the white dashed line correspond to scattering into the surface. The white cross represents the outgoing wavevector normal to the sample surface.}
    \label{fig:defect_scattering}
\end{figure}

To model defect-induced diffuse scattering, we use a supercell approach\cite{Staszewski2023BlueScattering}. For a pristine surface, only channels satisfying the diffraction condition \(\bm{K}_f=\bm{K}_i+\bm{G}\) carry non-zero intensity, while the remaining channels have zero intensity, left panel Fig.~\ref{fig:defect_scattering}. In the presence of a defect, translational symmetry is broken and intensity is redistributed away from the pristine Bragg channels into previously forbidden channels, right panel Fig.~\ref{fig:defect_scattering}. This redistributed intensity is taken as a measure of the defect-induced diffuse scattering.

The close-coupling calculations assume a frozen surface and therefore correspond to the \(\SI{0}{\kelvin}\) limit. To account for the finite sample temperatures used experimentally, we apply a semi-empirical correction to the calculated diffraction intensities. In the high-temperature limit, the thermal attenuation of elastic helium diffraction is described by
\begin{equation}
    I_{h\ell}(T)=I_{h\ell}(0)\,\mathrm{e}^{[-2W_{h\ell}(T)]},
    \;\;
    2W_{h\ell}(T)=C Q_{h\ell}^2 T,
    \label{eq:dw}
\end{equation}
where \(2W_{h\ell}(T)\) is the Debye--Waller (DW) exponent for diffraction channel \(h\ell\), \(Q_{h\ell}\) is the total momentum transfer for that channel, and \(C\) is a semi-empirical coefficient fixed by the measured temperature dependence of specular $\hkl(00)$ attenuation. Eq.\eqref{eq:dw} is the conventional high-temperature DW form for atom diffraction\cite{Farias1998AtomicSurfaces} and has previously been applied to MoS\textsubscript{2}\cite{Anemone2019ElectronPhononScattering}. Attenuation in the remaining channels is inferred from the measured specular attenuation using the \(Q^2\) scaling in Eq.~\ref{eq:dw}. Per-channel thermal correction is then applied to the calculated intensities.

CC simulations were performed for $5\times5$ supercells with varying defect density. The resulting intensities were fitted using the same lattice gas [eq. \eqref{eqn:lattice_gas}] and medium overlap [eq. (S1)] models as for the experimental data. The extracted scattering cross-sections are given in Table~\ref{tab:cross_sections}. They are significantly enhanced relative to the unit-cell area, but are approximately 25\% smaller than the experimental values. The calculated \hkl(-10) channel shows a channel-specific limitation for which neither model gave a satisfactory fit and is not used for quantitative validation.

\vspace{-0.5\baselineskip}

\begin{table}[H]
\begin{ruledtabular}
\begin{tabular}{c c c c c c }
Model & $\hkl(10)/\SI{}{\angstrom^2}$ & $\hkl(00)/\SI{}{\angstrom^2}$ & $\hkl(-20)/\SI{}{\angstrom^2}$ & $\hkl(-30)/\SI{}{\angstrom^2}$ \\
\hline
Lattice gas    & $18.2\pm0.2$ & $21.8\pm0.6$ & $15.4\pm0.5$ & $20.0\pm0.4$ \\
Medium overlap & $18.8\pm0.2$ & $22.4\pm0.7$ & $16.2\pm0.5$ & $20.7\pm0.4$ \\
\end{tabular}
\end{ruledtabular}
\caption{Effective He-defect cross-sections obtained by fitting the lattice gas [eq.\eqref{eqn:lattice_gas}] and medium overlap [eq. (S1)] models to the same CC-calculated diffraction intensities.}
\label{tab:cross_sections}
\end{table}

\clearpage
\bibliography{library.bib}


\end{document}


\preprint{prl}

\title{Supplemental Materials: Measuring vacancy-type defect density in monolayer semiconductors}

\author{Aleksandar Radic}\email{ar2071@cam.ac.uk}
\affiliation{Cavendish Laboratory, Department of Physics, University of Cambridge, JJ Thomson Ave, Cambridge, United Kingdom}%

\author{Nick von Jeinsen}%
\affiliation{Cavendish Laboratory, Department of Physics, University of Cambridge, JJ Thomson Ave, Cambridge, United Kingdom}%

\author{Vivian Perez}
\affiliation{Cavendish Laboratory, Department of Physics, University of Cambridge, JJ Thomson Ave, Cambridge, United Kingdom}%

\author{Ke Wang}
\affiliation{Cavendish Laboratory, Department of Physics, University of Cambridge, JJ Thomson Ave, Cambridge, United Kingdom}%

\author{Min Lin}
\affiliation{Cavendish Laboratory, Department of Physics, University of Cambridge, JJ Thomson Ave, Cambridge, United Kingdom}%

\author{Boyao Liu}
\affiliation{Cavendish Laboratory, Department of Physics, University of Cambridge, JJ Thomson Ave, Cambridge, United Kingdom}%

\author{Yiru Zhu}
\affiliation{Department of Materials Science and Metallurgy, University of Cambridge, 27 Charles Babbage Road, Cambridge, United Kingdom}%

\author{Ismail Sami}
\affiliation{Department of Materials Science and Metallurgy, University of Cambridge, 27 Charles Babbage Road, Cambridge, United Kingdom}%

\author{Kenji Watanabe}
\affiliation{Research Center for Electronic and Optical Materials, National Institute for Materials Science, 1-1 Namiki, Tsukuba 305-0044, Japan}

\author{Takashi Taniguchi}
\affiliation{Research Center for Materials Nanoarchitectonics, National Institute for Materials Science, 1-1 Namiki, Tsukuba 305-0044, Japan
}

\author{David Ward}
\affiliation{Cavendish Laboratory, Department of Physics, University of Cambridge, JJ Thomson Ave, Cambridge, United Kingdom}%

\author{Andrew Jardine}
\affiliation{Cavendish Laboratory, Department of Physics, University of Cambridge, JJ Thomson Ave, Cambridge, United Kingdom}%

\author{Akshay Rao}
\affiliation{Cavendish Laboratory, Department of Physics, University of Cambridge, JJ Thomson Ave, Cambridge, United Kingdom}%

\author{Manish Chhowalla}
\affiliation{Department of Materials Science and Metallurgy, University of Cambridge, 27 Charles Babbage Road, Cambridge, United Kingdom}%

\author{Sam Lambrick}
\affiliation{Cavendish Laboratory, Department of Physics, University of Cambridge, JJ Thomson Ave, Cambridge, United Kingdom}%
\affiliation{ISIS Facility, Rutherford Appleton Laboratory, Chilton, Didcot, Oxfordshire OX11 0QX, United Kingdom}

\date{\today}

\maketitle

\appendix

\renewcommand{\thefigure}{S\arabic{figure}}
\renewcommand{\thetable}{S\arabic{table}}
\renewcommand{\theequation}{S\arabic{equation}}
\setcounter{figure}{0}
\setcounter{equation}{0}
\setcounter{page}{1}

\onecolumngrid

\section{A comparison of current techniques used to measure defect density}

Table \ref{tab:defect_methods} provides a non-exhaustive comparison of the key characteristic and limits of the current standard techniques for vacancy-type defect density measurement (XPS = x-ray photoelectron spectroscopy, PL = photoluminescence, cAFM = conductive atomic force microscopy, STM = scanning tunnelling microscopy, (S)TEM = (scanning) transmission electron microscopy).

\onecolumngrid

\begin{table}[H]
\centering
\begin{tabularx}{\linewidth}{lXXXXXXX}

Technique & \makecell[l]{Days \\ /per sample} & Lab-based & Destructive & \makecell[l]{Sample \\requirements} & \makecell[l]{Lateral \\resolution} & Access time & Direct \\
\hline
\makecell[l]{Beamline XPS \\ (Diamond Light Source \\-- I06, I09, B07)} 
    & N/A & No & Yes & X, ultrahigh vacuum & 4$\times$20\,µm & months & Yes \\
\makecell[l]{}  &  &  &  &  &  & \\
\makecell[l]{Lab XPS \\ (ThermoFisher Nexsa G2)} 
    & 2 & Yes & Mode/material dependent & X, ultrahigh vacuum & 10--400\,µm & weeks & Yes \\
\makecell[l]{Low temp.\ PL \\ (Horiba Scientific)} 
    & 0.25 & Yes & Material dependent & Optically transparent & $\sim$1\,µm & weeks & No \\
\makecell[l]{cAFM \\ (Veeco Dimension Pro)} 
    & 2 & Yes & Mode/material dependent & Conductive, high vacuum & $<$1\,nm & weeks & \makecell[l]{Yes \\(statistical)}\\
STM 
    & 2 & Yes & Mode/material dependent & Conductive, high vacuum & $<$1\,nm & weeks & \makecell[l]{Yes \\(statistical)}\\
\makecell[l]{(S)TEM \\ (FEI Tecnai G2)} 
    & 2 & Yes & Yes & Conductive, ultrahigh vacuum & $<$1\,nm & weeks/months & \makecell[l]{Yes \\(statistical)}\\
\makecell[l]{Helium atom \\ micro-diffraction} 
    & 1 & Yes & No & High vacuum & $>$0.35\,µm & weeks & Yes \\
\hline
\end{tabularx}
\caption{A non-exhaustive list of standard characterisation techniques used to measure vacancy-type defect density in ML-TMDs. Time taken per sample includes estimates of typical sample preparation, measurement time and data analysis. Time to access refers to the typical availability of a given technique at the University of Cambridge, including research proposal submission/consideration if applicable. Costs for lab XPS, photoluminescence and helium atom micro-diffraction are taken as the full service mode rate for internal users as determined by the collaborative R\&D environment for physics at the Cavendish Laboratory (CORDE). Costs for cAFM and TEM taken as full service mode rate for internal users as determined by the Cambridge Centre for Gallium Nitride and Cambridge Advanced Imaging Centre, respectively. In all cases industry rates are higher and access times may be longer.}
\label{tab:defect_methods}
\end{table}
\clearpage

\twocolumngrid

\clearpage
\section{Sample information}

The native defect density of MoS\textsubscript{2} is reported as $\sim\SI{1e13}{\per\centi\metre\squared}$ whilst the upper limit is $\sim\SI{1e15}{\per\centi\metre\squared}$ before the lattice structure breaks down\cite{Yang2019SingleCatalysis}. Three samples of mechanically exfoliated monolayer MoS\textsubscript{2}, with increasing defect densities were prepared using high temperature annealing under a mixed Ar/H\textsubscript{2} (95\%/5\%) gas flow, as reported by Zhu \emph{et al.}\cite{Zhu2023Room-TemperatureDisulfide}. The bulk crystal used for exfoliation is the most commonly found 2H phase (2H-MoS\textsubscript{2}). To ensure precise knowledge of defect density we reproduce the annealing parameters exactly from the original work by Zhu \emph{et al.} \cite{Zhu2023Room-TemperatureDisulfide}. Exact sample annealing protocols and defect densities are listed in Table \ref{tab:defect_sample}.

\begin{table}[H]
\centering
\begin{tabular}{cccc}
\multirow{2}{*}{Sample} & \multicolumn{2}{c}{Annealing Protocol} & \multirow{2}{*}{\begin{tabular}[c]{@{}l@{}}Defect density \\ /$\times\SI{e14}{\per\centi\metre\squared}$\end{tabular}} \\
 & Time / $\SI{}{\hour}$ & Temp. / $\SI{}{\degreeCelsius}$ &  \\ \hline
1 & N/A & N/A & $\sim$0.1 (native) \\
2 & 0.5 & 550 & 0.62 \\
3 & 0.5 & 600 & 1.8\\\hline
\end{tabular}
\caption[Annealing parameters to induce controlled S-vacancy densities]{Annealing protocols used to induce S-vacancies in ML-MoS\textsubscript{2}: Ar/H\textsubscript{2} annealing protocol used to produce each ML- MoS\textsubscript{2} sample with changing S-vacancy defect density, as outlined by Zhu \emph{et al.}\cite{Zhu2023Room-TemperatureDisulfide}.}
\label{tab:defect_sample}
\end{table}

Samples were stored in a glovebox with argon atmosphere and transferred into a SHeM with no more than $\SI{1}{\hour}$ exposure to air. Owing to the chemically and electrically inert helium-4 probe sample preparation prior to insertion into the SHeM is minimal. The SiO\textsubscript{2}/Si substrate, onto which the MoS\textsubscript{2}/hBN flakes are deposited, are mounted to a custom SEM sample stub with sample heating\cite{ZhaoInstrumentationMicroscopy}. No additional treatments or surface coatings are required. The SHeM sample environment operates at high-vacuum ($\sim\SI{2e-8}{\milli\bar}$) pressures.  

\
\newpage
\section{Experimental procedure}

Each sample was heated to $\SI{220}{\degreeCelsius}$ for 2 hours to ensure the removal of surface contaminants from the MoS\textsubscript{2}, which helium has a high sensitivity to. Prior to heating, diffractive measurements of the samples were attempted and yielded disordered scattering rather than ordered diffraction, as one would expect from a surface covered in disordered adsorbates. Helium has a large scattering cross-section for physisorbed (van der Waals bound) surface adsorbates, due to the low energy of the atoms and a strong attractive component in its interaction potential. Therefore, even low coverages of surface adsorbates will create predominantly disordered scattering. After heating, the same measurements were repeated and showed the expected diffraction corresponding to the trigonal arrangement of the top sulphur atoms. All subsequent defect density measurements were taken at a temperature of $\SI{120}{\degreeCelsius}$ to ensure that vacuum contaminants would not re-adsorb to the sample. 

To quantify defect density, SHeM diffraction measurements were performed on each sample and the ratio of detected helium intensity which exhibits ordered vs. disordered scattering used to determine the degree of order present on the sample surface. In figure 2, the diffraction patterns of the two extrema of surface order are shown. The pristine bulk MoS\textsubscript{2} and pristine monolayer MoS\textsubscript{2}/hBN/SiO\textsubscript{2} represent the maximum `order' achievable in the presented samples, and conversely, the pristine monolayer MoS\textsubscript{2}/SiO\textsubscript{2} represents the minimum (via substrate interaction). Due to real-world experimental constraints, there will always be some disordered background signal detected arising from multiple scattering\cite{Lambrick2020MultipleMicroscopy,Lambrick2018AImaging} and imperfect samples (e.g. $>\SI{0}{\per\centi\metre\squared}$ defect density). By including bulk MoS\textsubscript{2}/SiO\textsubscript{2} and exposed SiO\textsubscript{2} on each sample one can perform normalisation measurements to determine the amount of disordered scattering background signal to subtract it from diffraction measurements of the defective sample.

\section{Consideration of overlapping He-defect scattering cross-sections}
\begin{figure}[H]
    \centering
    \includegraphics[width=85mm]{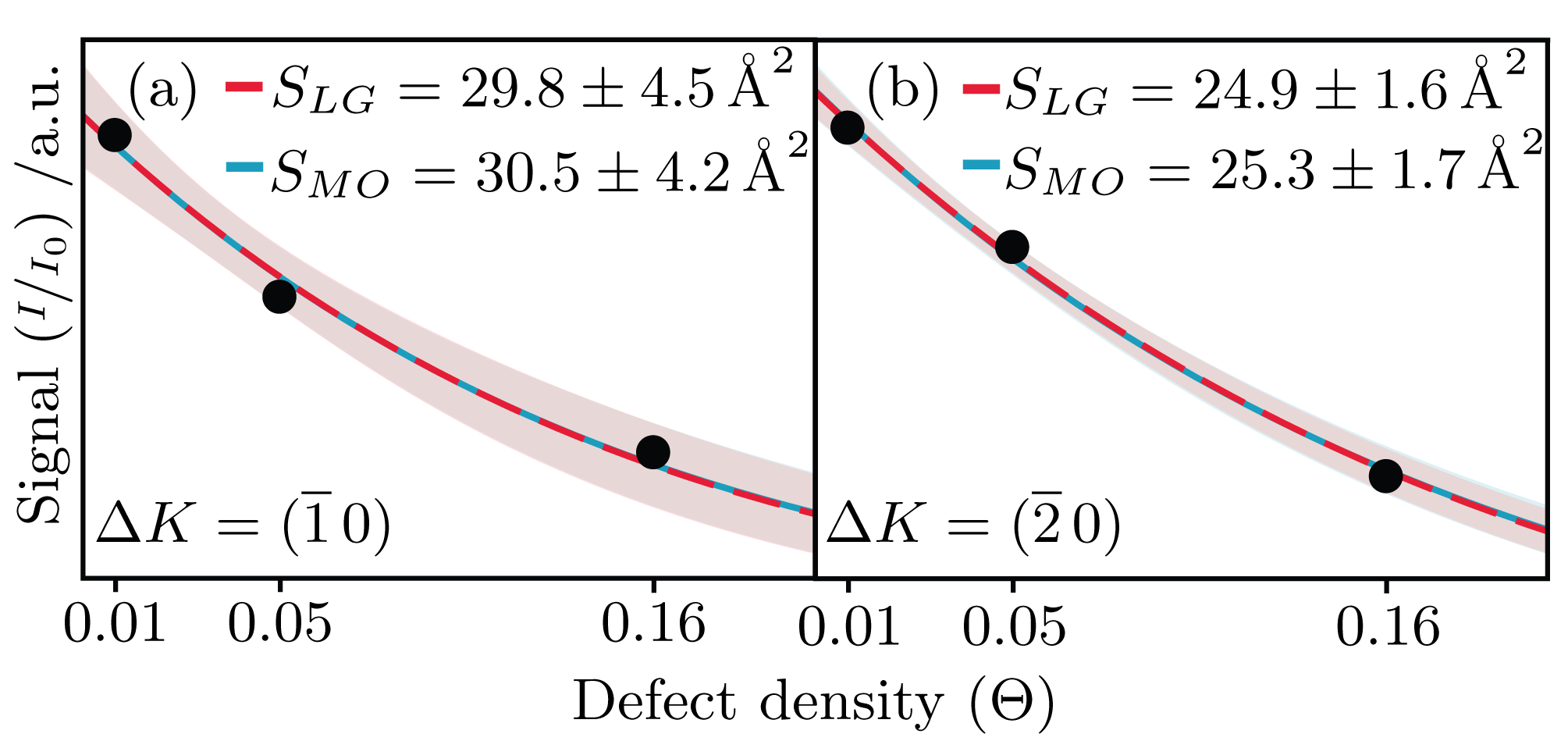}
    \caption{Comparison of the lattice gas (eq. (1)) and `medium overlap' (eq. \ref{eq:medium}) models, the former assumes that defect cross-section are isolated, the latter permits overlap. We find excellent agreement between the two models for both \hkl(-10) (a) and \hkl(-20) (b) and therefore recommend the lattice gas model for its simplicity and applicability to all lattice geometries.}
    \label{fig:model_comparison}
\end{figure}

The lattice gas equation in eq. (1) assumes cross-sections significantly larger than the unit cell area and negligible overlap. In the case of moderate cross sections and moderate overlap alternative expressions can be used, for an hexagonal lattice such as MoS\textsubscript{2}\cite{Poelsema1989ScatteringSurfaces}
\begin{equation}\label{eq:medium}
    \begin{split}
        \frac{A}{A_0}&=1 - \Theta\eta^2 + \Theta^2(7\eta^2 - 16\eta + 10) \\ 
        &- \Theta^3(10\eta^2 - 28\eta + 20) + \Theta^4(4\eta^2 - 12\eta + 9)
    \end{split}
\end{equation}
where $A/A_0$ is the reduction in amplitude, $S$ is the quantum cross section and $\eta=\sqrt{\nicefrac{S}{A_\mathrm{cell}}}$. Equation \ref{eq:medium} is valid for $1.5\leq\eta\leq2$. Similar expressions may be derived for different ranges of $\eta$ and for different surface geometry. In Figure \ref{fig:model_comparison} we compare the Eqs. (1) and \ref{eq:medium} and find excellent agreement in the fits and resulting scattering cross-sections. All fits were performed with \texttt{lmfit} to perform non-linear least-squares\cite{NewvilleLMFITPython}.

\section{DFT details}

All DFT calculations were performed using CASTEP \cite{Clark2005FirstCASTEP}, a plane-wave-based DFT code employing periodic boundary conditions. The exchange-correlation energy was treated using the Perdew–Burke–Ernzerhof (PBE) \cite{Perdew1996GeneralizedSimple} functional within the generalised gradient approximation, with DFT-D3 \cite{Moellmann2014DFT-D3Crystals} dispersion corrections included to account for van der Waals interactions. A kinetic energy cutoff of 600 eV was used for the plane-wave basis set, and ultrasoft pseudopotentials \cite{Vanderbilt1990SoftFormalism} were employed for all elements. To eliminate spurious interactions between periodic images, a vacuum spacing of $\SI{25}{\angstrom}$ was introduced along the out-of-plane direction. A $4\times4\times1$ supercell of MoS\textsubscript{2} was used to prevent artificial interactions between adjacent He atoms. The interaction energy between a single helium atom and the MoS\textsubscript{2} surface was evaluated under the rigid-surface approximation, where the helium atom is assumed not to significantly perturb the substrate structure. Accordingly, the MoS\textsubscript{2} slab was fully relaxed first, and the He atom was subsequently placed at selected sites above the surface for single-point total energy calculations. A $4\times4\times1$ Monkhorst–Pack \cite{Monkhorst1976SpecialIntegrations} k-point mesh was used. Electronic energy convergence was set to $\SI{1e-8}{\electronvolt}$, and a force convergence criterion of $\SI{0.03}{\electronvolt\per\angstrom}$ was applied during structural relaxations.

\newpage

\section{Measuring the lattice constant}

\begin{figure}[H]
    \centering
    \includegraphics[width=1\linewidth]{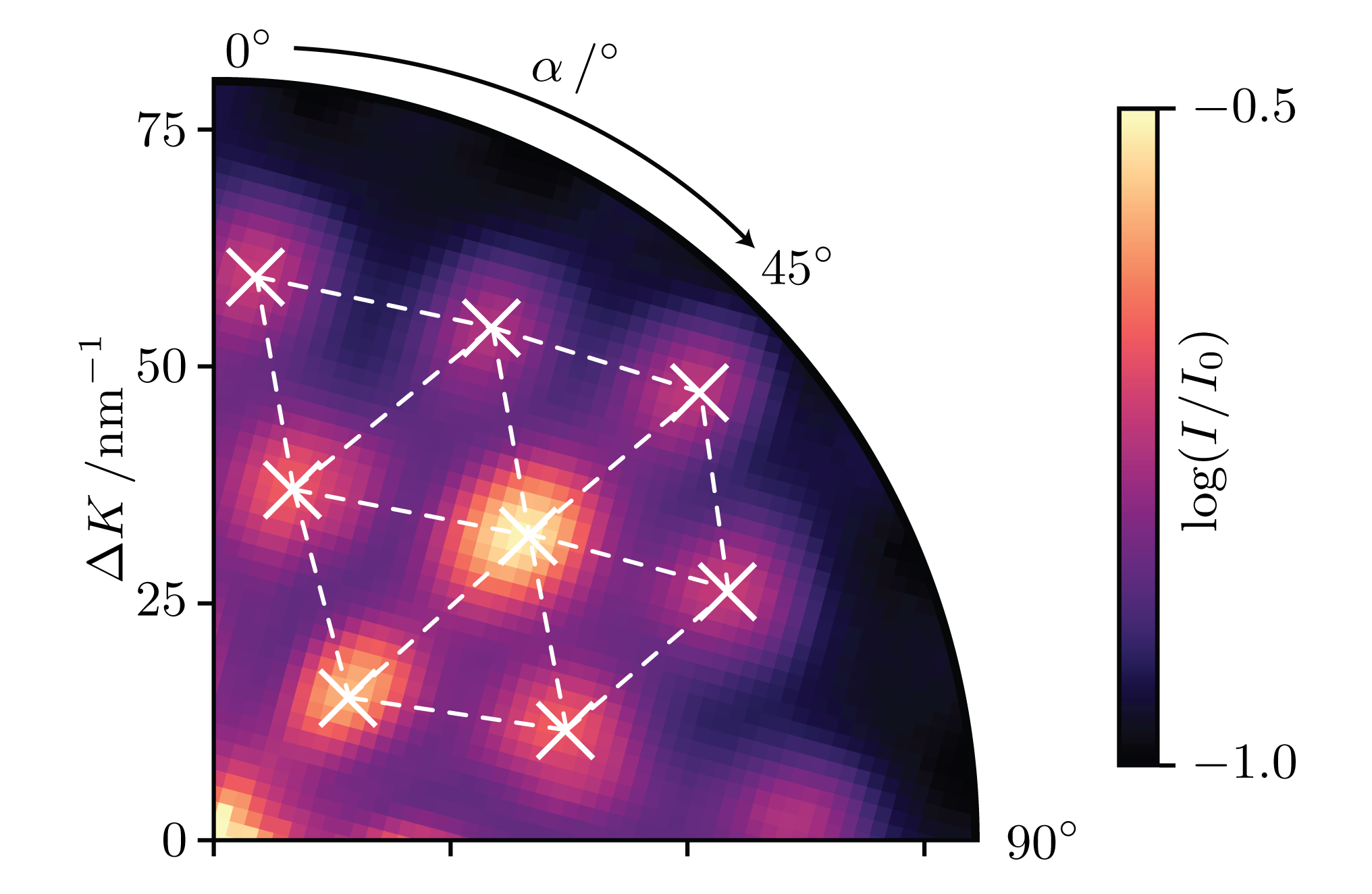}
    \caption{Plot of 2D Bragg diffraction for native defect density monolayer MoS\textsubscript{2} with the principal azimuth (\hkl<10>) lying along $\alpha=\SI{45}{\degree}$ and measured at $\SI{373}{\kelvin}$. The average of distances (dashed lines) between adjacent peaks (crosses) is then used to calculate the lattice constant, $a=3.20\pm\SI{0.07}{\angstrom}$. Literature values are $\sim\SI{3.16}{\angstrom}$\cite{Nguyen2019FabricationLIBs}. The centres of diffraction peaks were identified using 2D Gaussian fitting.}
    \label{fig:lattice_constant}
\end{figure}

Figure \ref{fig:lattice_constant} shows the raw 2D diffraction data acquired from the monolayer MoS\textsubscript{2}. Diffraction peak locations were identified (with the exception of specular which is at the centre of the plot) by fitting a 2D Gaussian and 2nd order polynomial background to each peak – the same model as used by von Jeinsen \emph{et al.}\cite{vonJeinsen20232DSpot}. In figure \ref{fig:lattice_constant} the fitted peak locations are shown with crosses and the nearest-neighbour peak spacings with dashed lines. All inter-peak spacings were averaged, with the standard error being calculated for that average (the number of independent measurements was taken to be the number of diffraction peaks used).

\newpage
\section{Temperature dependence of helium atom micro-diffraction}

\begin{figure}[H]
    \centering
    \includegraphics[width=\linewidth]{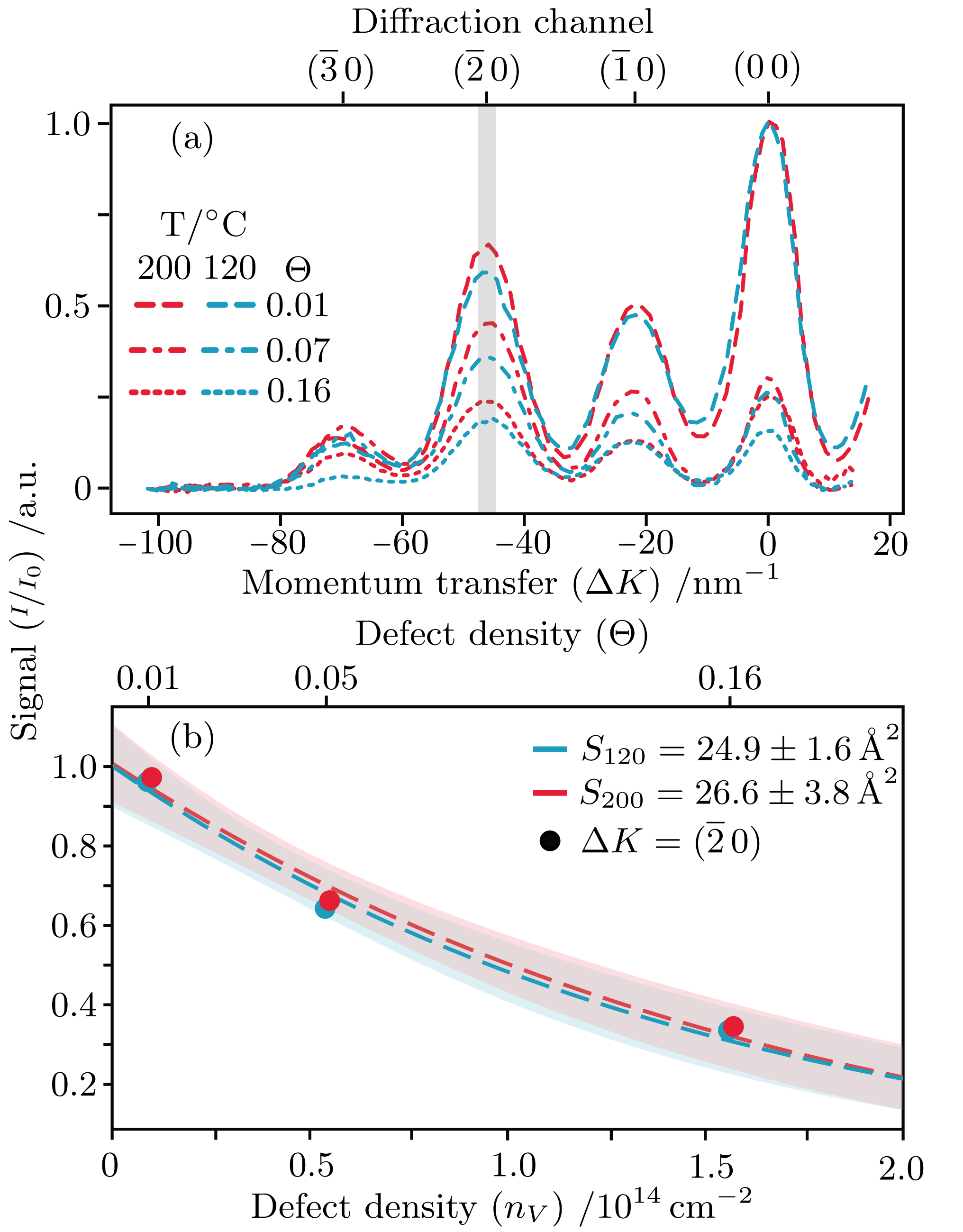}
    \caption{Panel (a), line scans of increasing defect density monolayer MoS\textsubscript{2} showing a consistent decrease in diffracted intensity as defect density increases at $\SI{120}{\degreeCelsius}$ and $\SI{200}{\degreeCelsius}$. $\SI{120}{\degreeCelsius}$ is the same as in figure 3 (main text). Panel (b) shows intensities extracted using Gaussian fitting to the $\hkl(-20)$ diffraction peaks at both temperatures and strong agreement with the lattice gas model from equation 1 in both cases. Solely considering the effect of temperature on Bragg diffraction, according to Debye-Waller attenuation, the lower temperature should have increased intensity in ordered scattering, thereby decreasing experimental errors. This neglects the possibility that vacuum contaminants will also adsorb to the surface more readily at lower temperatures, reducing the degree of order and introducing cosine-like (diffuse) rather than Bragg-like scattering.}
    \label{fig:200_vs_120}
\end{figure}

\newpage
\section{Fitting the lattice gas equation at different diffraction conditions}

\begin{figure}[H]
    \centering
    \includegraphics[width=\linewidth]{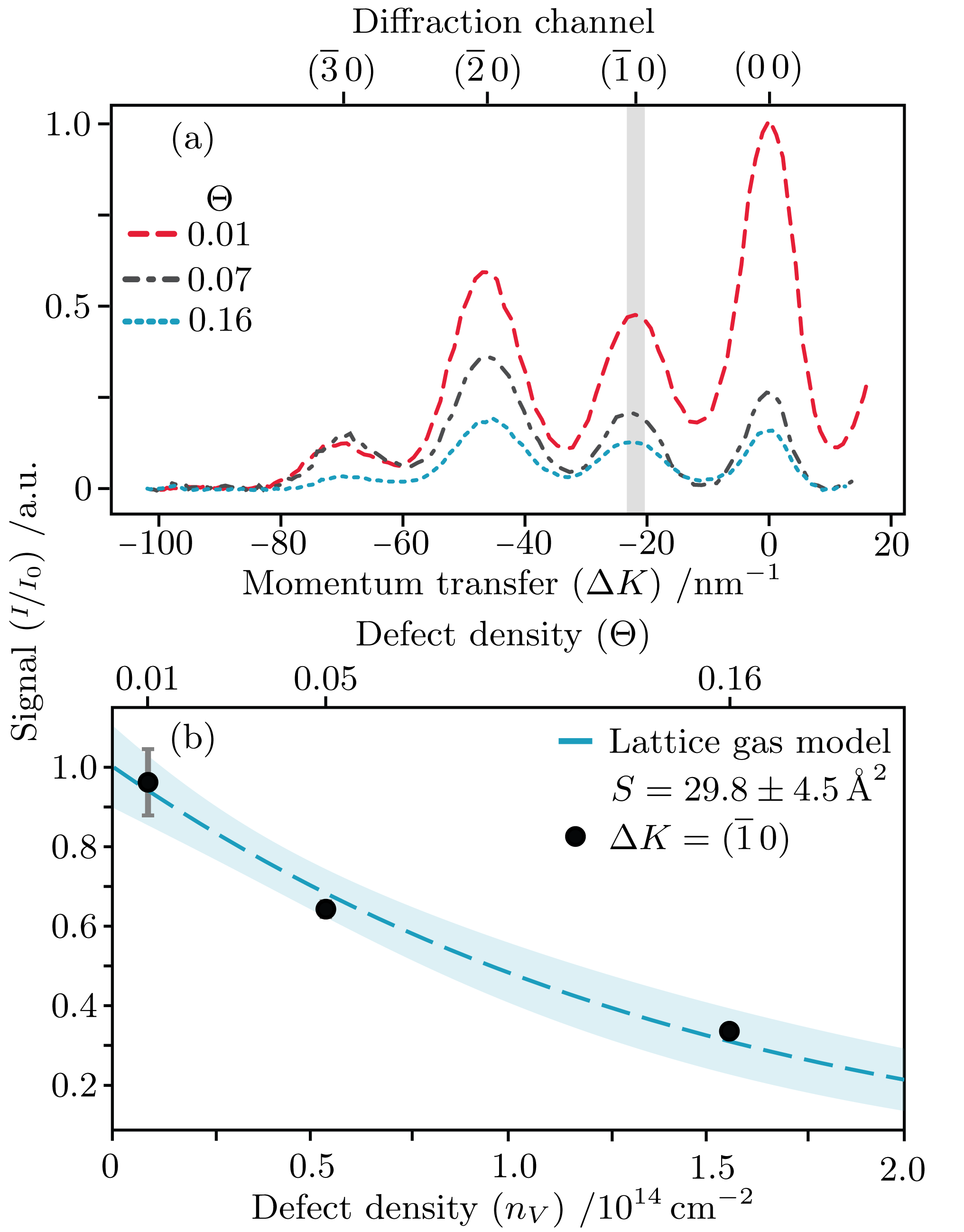}
    \caption{Panel (a), line scans of increasing defect density monolayer MoS\textsubscript{2} acquired at $\SI{120}{\degreeCelsius}$ showing a consistent decrease in diffracted intensity as defect density increases, as previously shown in figure 3 (main text). Panel (b) shows intensities extracted using Gaussian fitting to the $\hkl(-10)$ diffraction peaks and qualitative agreement with the lattice gas model from equation 1. The $\hkl(-10)$ peak suffers from lower intensities and an increased sensitivity to sample tilt compared to $\hkl(-20)$, which negatively effects the quality of the fit.}
    \label{fig:lattice_gas_10}
\end{figure}

\newpage
\section{Details on close-coupled scattering calculations}

The close-coupling (CC) method is a quantum mechanical approach used to calculate the scattering of a helium matter wave from a surface potential\cite{ManolopoulosIterativeApproach,Riley2008AnalysisResonances,Avidor2016HeliumSurfaces,Staszewski2023BlueScattering}. In this framework, the helium atom is described by a wavefunction that satisfies the time-independent Schr\"odinger equation in the presence of a three-dimensional potential V(x,y,z), obtained from density functional theory (DFT).

Owing to the periodicity of the surface, the wavefunction is expanded in a basis of plane waves corresponding to reciprocal lattice vectors $G_{h\ell}$, such that each basis function represents a diffraction channel with parallel momentum $\mathbf{K}_f=\mathbf{K}_i+\mathbf{G}_{h\ell}$. Substitution into the Schr\"odinger equation yields a set of coupled differential equations for the channel amplitudes, which are solved numerically using the MultiScat package\cite{CambridgeAtomScatteringCentre2020GitHubCambridge-Atom-Scattering-Centre/Multiscat}.

The solution provides the scattering amplitudes for all outgoing channels, from which the diffraction intensities $I(\mathbf{K}_f,\mathbf{K}_i)$ are obtained. For a pristine surface, only kinematically allowed channels ($\mathbf{K}_f=\mathbf{K}_i+\mathbf{G}$) carry intensity, whereas in the presence of defects, symmetry breaking redistributes intensity into previously forbidden channels, which we take as a measure of the defect-induced diffuse background. The use of periodically repeated disordered supercells follows the approach of Ref.\cite{Avidor2016HeliumSurfaces}, while the resulting redistribution from diffraction into diffuse scattering is closely related to Ref.\cite{Staszewski2023BlueScattering}.

In this work, the CC calculations are used to compute diffraction intensities for surfaces with varying defect densities, constructed using supercells. The resulting intensities are analysed using the same lattice gas model as applied to the experimental data, allowing extraction of effective helium–defect scattering cross-sections.

The CC calculations assume a static (frozen) surface potential and therefore do not explicitly include thermal motion or inelastic scattering processes. These effects are incorporated separately via an empirical Debye–Waller correction in the following section, which rescales the calculated intensities according to the experimentally measured thermal attenuation.

\subsection{Empirical Debye--Waller correction to the elastic diffraction intensities}

The close-coupled calculations employed here describe scattering from a static surface and therefore do not account for the thermal attenuation of the elastic diffraction intensities. To incorporate this effect empirically, we rescale each calculated diffraction intensity using the experimentally measured Debye--Waller attenuation of the specular beam.

For elastic atom scattering, thermal vibrations reduce the elastic intensity according to
\begin{equation*}
    I_{h\ell}(T)=I_{h\ell,0}\exp[-2W_{h\ell}(T)],
    \label{eqn:dw_definition}
\end{equation*}
where \(I_{h\ell,0}\) is the elastic intensity in the frozen-lattice limit and \(2W_{h\ell}(T)\) is the Debye--Waller exponent for diffraction channel \((h,\ell)\). We use the high-temperature form for Debye-Waller attenuation, as defined by Farias and Rieder\cite{Farias1998AtomicSurfaces}, as
\begin{equation*}
    2W_{h\ell}(T)=C Q_{h\ell}^2 T,
    \label{eq:scaling_relation}
\end{equation*}
where \(Q_{h\ell}\) is the total momentum transfer for channel \((h,\ell)\), and \(C\) is a semi-empirical coefficient fixed by the Debye-Waller attenuation of the specular $\hkl(00)$ channel. Channels with larger momentum transfer therefore undergo stronger thermal attenuation.

Using the measured attenuation of the specular beam, the constant \(CT\) can be eliminated to give
\begin{equation*}
    I_{h\ell}(T)
    = I_{h\ell,0}
    \left(
    \frac{I_{00}(T)}{I_{00,0}}
    \right)^{Q_{h\ell}^2/Q_{00}^2}.
    \label{eqn:deriv_dw}
\end{equation*}

Equivalently, writing the measured specular Debye--Waller exponent as
\begin{equation*}
    2W_{00}(T)=\left(\SI{2.98(0.17)e-3}{\per\kelvin}\right)T,
\end{equation*}
gives
\begin{equation}
    I_{h\ell}(T)=I_{h\ell,0}\exp\!\left[-2W_{00}(T)
    \left(\frac{Q_{h\ell}^2}{Q_{00}^2}\right)\right].
    \label{eq:DW_final}
\end{equation}

\begin{figure}[H]
    \centering
    \includegraphics[width=85mm]{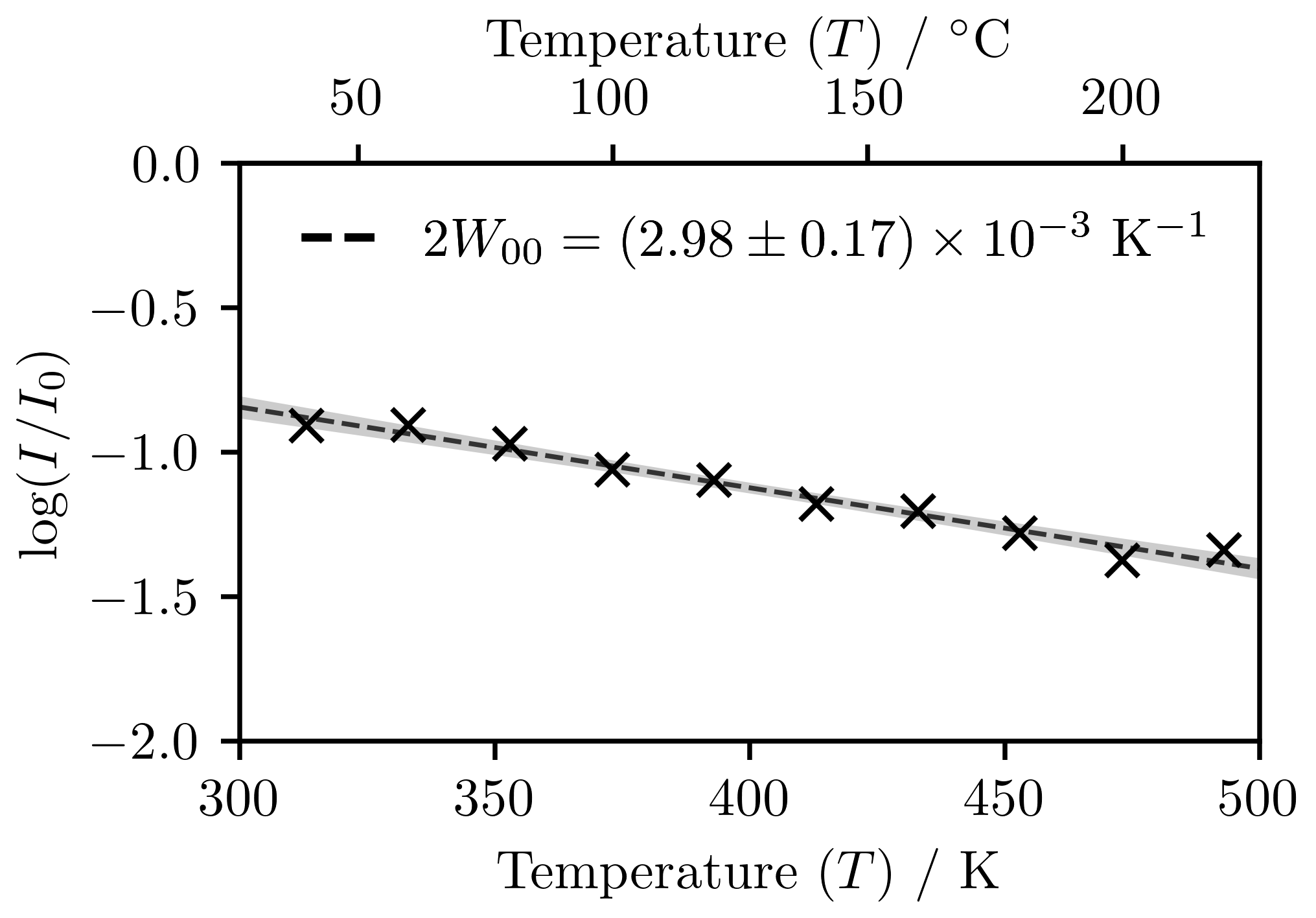}
    \caption{Experimentally measured specular Debye--Waller exponent, \(2W_{00}(T)=\left(\SI{2.98(0.17)e-3}{\per\kelvin}\right)T\), for the lowest defect density sample used in the present work (\(n_V=\SI{0.1e14}{\per\centi\metre\squared}\), \(\Theta_1=0.01\)). This measured attenuation is used to apply the empirical Debye--Waller correction to the close-coupled scattering calculations according to Eq.~\ref{eq:DW_final}.}
    \label{fig:DW_specular}
\end{figure}

This empirical procedure preserves the experimentally measured thermal attenuation in the specular channel while extending it to off-specular diffraction peaks through the standard Debye--Waller scaling with momentum transfer. In this way, thermal effects can be incorporated into the close-coupled scattering calculations without introducing additional microscopic assumptions regarding the surface phonon spectrum.

\subsection{Derivation of the total momentum transfer $Q_{h\ell}^2$}

We begin from the definition of the momentum transfer for elastic scattering:
\begin{equation*}
    \mathbf{Q}_{h\ell} = \mathbf{k}_f - \mathbf{k}_i,
\end{equation*}
and therefore
\begin{equation*}
    Q_{h\ell}^2 = |\mathbf{k}_f - \mathbf{k}_i|^2.
\end{equation*}

\paragraph{Scattering geometry.}
We define a coordinate system in which \(z\) is the surface normal, \(x\) lies in the scattering plane, and \(y\) is the in-plane direction perpendicular to the scattering plane.

The incident wavevector lies in the \(x\)--\(z\) plane and is given by
\begin{equation*}
    \mathbf{k}_i = (k\sin\theta_i,\, 0,\,-k\cos\theta_i),
\end{equation*}
where \(\theta_i\) is the angle of incidence and \(k = |\mathbf{k}_i|\).

\paragraph{Parallel momentum conservation.}

For diffraction from a periodic surface, the parallel component of momentum is conserved modulo a reciprocal lattice vector \(\mathbf{G}_{h\ell}\):
\begin{equation*}
    \mathbf{k}_{\parallel,f} = \mathbf{k}_{\parallel,i} + \mathbf{G}_{h\ell}.
\end{equation*}

Writing
\begin{equation*}
    \mathbf{G}_{h\ell} = (G_x, G_y, 0),
\end{equation*}
the final wavevector becomes
\begin{equation*}
    \mathbf{k}_f = (k\sin\theta_i + G_x,\; G_y,\; k_{z,f}).
\end{equation*}

For elastic scattering, energy is conserved, so
\begin{equation*}
    |\mathbf{k}_f| = |\mathbf{k}_i| = k,
\end{equation*}
which gives
\begin{equation*}
    (k\sin\theta_i + G_x)^2 + G_y^2 + k_{z,f}^2 = k^2.
\end{equation*}

Rearranging gives
\begin{equation*}
    k_{z,f} = \sqrt{k^2 - (k\sin\theta_i + G_x)^2 - G_y^2}.
\end{equation*}

The momentum transfer vector is therefore
\begin{align*}
    \mathbf{Q}_{h\ell}
    &= \mathbf{k}_f - \mathbf{k}_i \\
    &= \bigl((k\sin\theta_i + G_x) - k\sin\theta_i,\;
             G_y,\;
             k_{z,f} - (-k\cos\theta_i)\bigr) \\
    &= (G_x,\; G_y,\; k_{z,f} + k\cos\theta_i).
\end{align*}

Taking the squared magnitude,
\begin{equation*}
    Q_{h\ell}^2 = G_x^2 + G_y^2 + (k_{z,f} + k\cos\theta_i)^2.
\end{equation*}

Substituting for \(k_{z,f}\) gives
\begin{equation}
Q_{h\ell}^2 =
G_x^2 + G_y^2 +
\left(
    \sqrt{k^2 - (k\sin\theta_i + G_x)^2 - G_y^2}
    + k\cos\theta_i
\right)^2.
\label{eqn:Qhl2}
\end{equation}

For the specular condition (\(G_x = G_y = 0\)), this reduces to
\begin{equation*}
    Q_{00}^2 = (2k\cos\theta_i)^2 = 4k_{z,i}^2,
\end{equation*}
as expected.

\begin{figure}[H]
    \centering
    \includegraphics[width=70mm]{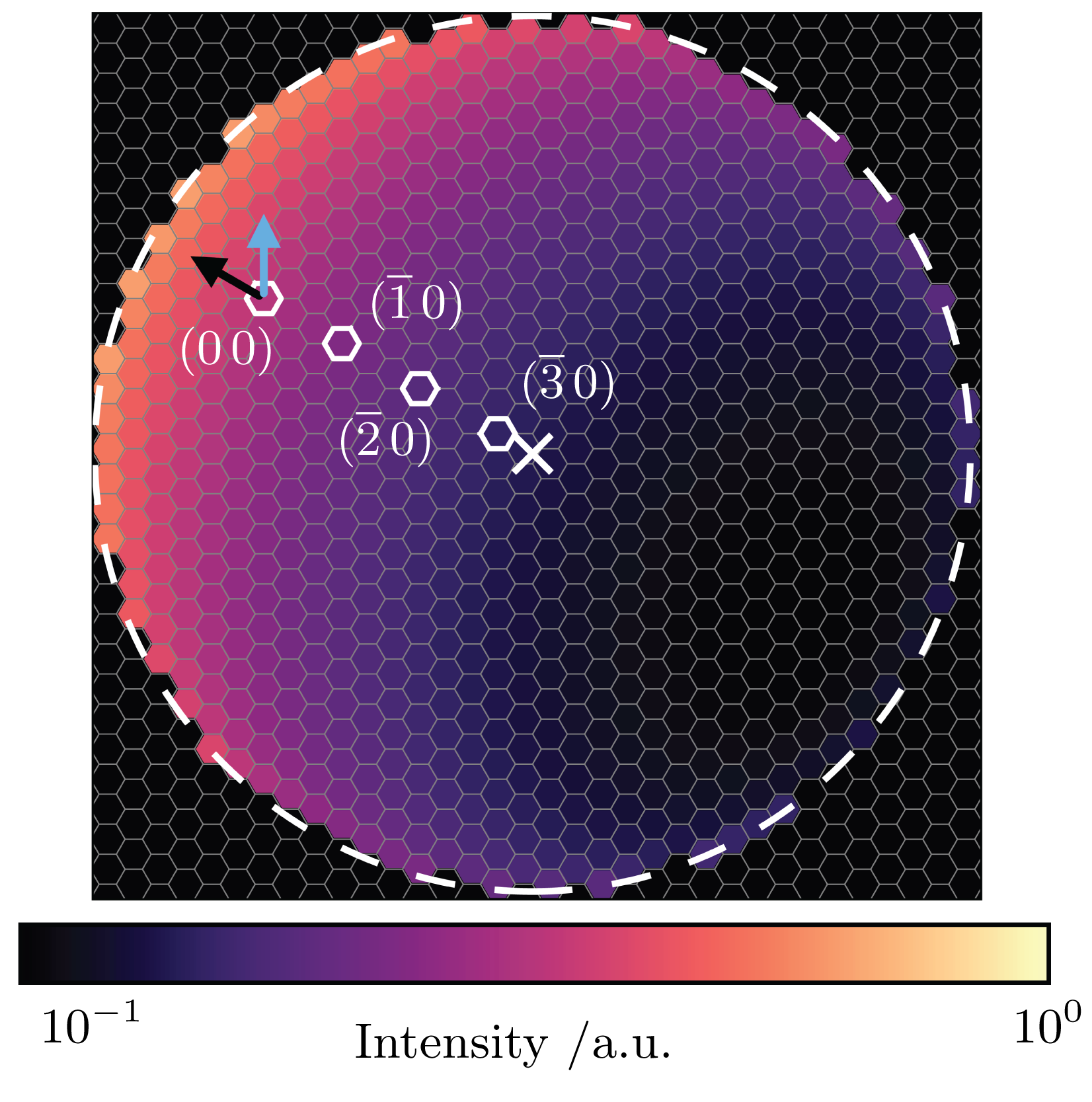}
    \caption{Thermal attenuation factor $I_{h\ell}(T)/I_{h\ell,0}$ for diffraction channels in $(h,\ell)$, calculated from Eq.~\eqref{eq:DW_final} at $T=\SI{120}{\degreeCelsius}$ to match the experimental results shown in Figure 3 (main text). Each cell represents the multiplicative thermal correction applied to the calculated intensities and are shown independently of the underlying close-coupled results. Increasing total momentum transfer $Q_{h\ell}$ leads to stronger attenuation for higher-order diffraction peaks.}
    \label{fig:thermal_factor}
\end{figure}

In Figure \ref{fig:thermal_factor} we present the empirical correction factor for all diffraction channels evaluated at sample temperature $T=\SI{120}{\degreeCelsius}$ using Eqn.\ref{eq:DW_final} and the experimentally measured Debye-Waller factor for the specular peak of the lowest defect density sample used in the current work ($2W_{00}=\SI{2.98(0.17)e-3}{\per\kelvin}$).

The resulting empirical attenuation factors at \(T=\SI{120}{\degreeCelsius}\) for the experimentally measured diffraction channels are
\begin{align*}
\frac{I_{\hkl(0,0)}(T)}{I_{\hkl(0,0),0}} &= 0.310, &
\frac{I_{\hkl(-1,0)}(T)}{I_{\hkl(-1,0),0}} &= 0.229, \\
\frac{I_{\hkl(-2,0)}(T)}{I_{\hkl(-2,0),0}} &= 0.179, &
\frac{I_{\hkl(-3,0)}(T)}{I_{\hkl(-3,0),0}} &= 0.146.
\end{align*}

\clearpage

\bibliography{library.bib}